\documentclass[a4paper,12pt]{article}
\pdfoutput=1

\usepackage{graphicx}
\usepackage{amsmath,amssymb,amsthm,mathtools}
\usepackage{rotating}
\usepackage{color}
\usepackage{xcolor}

\usepackage{caption}
\usepackage{subcaption}
\usepackage{float}
\usepackage[toc,page]{appendix}
\usepackage[numbers,sort&compress]{natbib}
\allowdisplaybreaks

\makeatletter
 
\def\paragraph{\@startsection{paragraph}{4}{\z@}{+2.00ex plus
 +1ex minus +.2ex}{1.5ex plus .2ex}{\it\normalsize}}
 
\def\section{\@startsection {section}{1}{\z@}{+3.0ex plus +1ex minus
  +.2ex}{2.3ex plus .2ex}{\normalsize\bf}}
\def\subsection{\@startsection{subsection}{2}{\z@}{+2.5ex plus +1ex
minus +.2ex}{1.5ex plus .2ex}{\normalsize\bf}}
\def\subsubsection{\@startsection{subsubsection}{3}{\z@}{+3.25ex plus
 +1ex minus +.2ex}{1.5ex plus .2ex}{\normalsize\bf}}

\@addtoreset{equation}{section}

\def\appendix{\par
 \setcounter{section}{0} 
 \setcounter{subsection}{0}
 \setcounter{equation}{0}
 \def\thesection{\Alph{section}}}
 
\expandafter\ifx\csname mathrm\endcsname\relax\def\mathrm#1{{\rm #1}}\fi
 
\makeatletter

\newcount\@tempcntc
\def\@citex[#1]#2{\if@filesw\immediate\write\@auxout{\string\citation{#2}}\fi
  \@tempcnta\z@\@tempcntb\m@ne\def\@citea{}\@cite{\@for\@citeb:=#2\do
    {\@ifundefined
       {b@\@citeb}{\@citeo\@tempcntb\m@ne\@citea
        \def\@citea{,\penalty\@m\ }{\bf ?}\@warning
       {Citation `\@citeb' on page \thepage \space undefined}}%
    {\setbox\z@\hbox{\global\@tempcntc0\csname
b@\@citeb\endcsname\relax}%
     \ifnum\@tempcntc=\z@ \@citeo\@tempcntb\m@ne
       \@citea\def\@citea{,\penalty\@m}
       \hbox{\csname b@\@citeb\endcsname}%
     \else
      \advance\@tempcntb\@ne
      \ifnum\@tempcntb=\@tempcntc
      \else\advance\@tempcntb\m@ne\@citeo
      \@tempcnta\@tempcntc\@tempcntb\@tempcntc\fi\fi}}\@citeo}{#1}}

\def\@citeo{\ifnum\@tempcnta>\@tempcntb\else\@citea
  \def\@citea{,\penalty\@m}%
  \ifnum\@tempcnta=\@tempcntb\the\@tempcnta\else
   {\advance\@tempcnta\@ne\ifnum\@tempcnta=\@tempcntb \else
\def\@citea{--}\fi
    \advance\@tempcnta\m@ne\the\@tempcnta\@citea\the\@tempcntb}\fi\fi}

\def\asymp#1%
{\mathrel{\raisebox{-.4em}{$\widetilde{\scriptstyle #1}$}}}

\def\Nequal#1%
{\mathrel{\raisebox{-.5em}{$\stackrel{=}{\scriptstyle\rm#1}$}}}
\def\dsl{\mathpalette\make@slash}
\def\make@slash#1#2{\setbox\z@\hbox{$#1#2$}%
  \hbox to 0pt{\hss$#1/$\hss\kern-\wd0}\box0}

\def\beq#1\eeq{\begin{equation}#1\end{equation}}
\def\beqar{\begin{eqnarray}}
\def\eeqar{\end{eqnarray}}
\def\barr#1{\begin{array}{#1}}
\def\earr{\end{array}}
\def\bfi{\begin{figure}}
\def\efi{\end{figure}}
\def\btab{\begin{table}}
\def\etab{\end{table}}
\def\bce{\begin{center}}
\def\ece{\end{center}}
\def\nn{\nonumber}

\def\Ga{\Gamma}

\def\refeq#1{\mbox{(\ref{#1})}}

\def\reffi#1{\mbox{Fig.~\ref{#1}}}

\def\refta#1{\mbox{Tab.~\ref{#1}}}

\def\refse#1{\mbox{Section~\ref{#1}}}

\def\refapp#1{\mbox{App.~\ref{#1}}}
\def\citere#1{\mbox{Ref.~\cite{#1}}}
\def\citeres#1{\mbox{Refs.~\cite{#1}}}
 
\newcommand{\TeV}{\unskip\,\mathrm{TeV}}
\newcommand{\GeV}{\unskip\,\mathrm{GeV}}
\newcommand{\MeV}{\unskip\,\mathrm{MeV}}

\newcommand{\ri}{{\mathrm{i}}}
\newcommand{\re}{{\mathrm{e}}}

\newcommand{\rw}{{\mathrm{w}}}
\newcommand{\rB}{{\mathrm{B}}}
\newcommand{\rF}{{\mathrm{F}}}

\newcommand{\rR}{{\mathrm{R}}}

\newcommand{\rT}{{\mathrm{T}}}

\newcommand{\M}{{\cal{M}}}

\def\mathswitchr#1{\relax\ifmmode{\mathrm{#1}}\else$\mathrm{#1}$\fi}

\newcommand{\Pq}{\mathswitch q}
\newcommand{\PW}{\mathswitchr W}
\newcommand{\Pw}{\mathswitchr w}
\newcommand{\PZ}{\mathswitchr Z}
\newcommand{\PA}{\mathswitchr A}

\newcommand{\PH}{\mathswitchr H}

\newcommand{\Ps}{\mathswitchr s}

\newcommand{\Pb}{\mathswitchr b}
\newcommand{\Pt}{\mathswitchr t}
\newcommand{\PPA}{\mathswitchr P\mathswitchr A}

\newcommand{\Pl}{\ell}
 
\def\mathswitch#1{\relax\ifmmode#1\else$#1$\fi}

\newcommand{\MW}{\mathswitch {M_\PW}}

\newcommand{\MZ}{\mathswitch {M_\PZ}}

\newcommand{\Mt}{\mathswitch {m_\Pt}}

\newcommand{\sw}{\mathswitch {s_\Pw}}
\newcommand{\cw}{\mathswitch {c_\Pw}}

\def\Re{\mathop{\mathrm{Re}}\nolimits}

\newcommand{\OS}{\mathrm{OS}}

\newcommand{\LO}{\mathrm{LO}}
\newcommand{\NLO}{\mathrm{NLO}}
\newcommand{\NNLO}{\mathrm{NNLO}}
\newcommand{\QCD}{\mathrm{QCD}}

\newcommand{\z}{\setbox0\hbox{+}\hbox to \wd0{\hss0\hss}}

\def\limfunc#1{\mathop{\rm #1}}

\def\Re{\limfunc{Re}}

\def\slash#1{{\setbox0=\hbox{$#1$}
  \rlap{\ifdim\wd0>.7em\kern.22\wd0\else\kern.1\wd0\fi /}#1}}

\def\braket#1#2{\left\langle #1\vphantom{#2}
  \right. \kern-2.5pt\left| #2\vphantom{#1}\right\rangle }

\def\M{{\cal M}}
\def\O{{\cal O}}

\def\rT{{\mathrm{T}}}

\def\CF{C_{\mathrm{F}}}

\def\alphas{\alpha_{\mathrm{s}}}

\newcommand{\KIRA}{{\sc KIRA}}

\def\draftdate{\relax}
\def\mda{\relax}
\def\mua{\relax}
\def\mla{\relax}
\def\Mda{\relax}
\def\Mua{\relax}
\def\Mla{\relax}
\def\draft{
\def\thtystars{******************************}
\def\sixtystars{\thtystars\thtystars}
\typeout{}
\typeout{\sixtystars**}
\typeout{* Draft mode!
         For final version remove \protect\draft\space in source file *}
\typeout{\sixtystars**}
\typeout{}
\def\draftdate{\today}
\def\mua{\marginpar[\boldmath\hfil$\uparrow$]%
                   {\boldmath$\uparrow$\hfil}%
                    \typeout{marginpar: $\uparrow$}\ignorespaces}
\def\mda{\marginpar[\boldmath\hfil$\downarrow$]%
                   {\boldmath$\downarrow$\hfil}%
                    \typeout{marginpar: $\downarrow$}\ignorespaces}
\def\mla{\marginpar[\boldmath\hfil$\rightarrow$]%
                   {\boldmath$\leftarrow $\hfil}%
                    \typeout{marginpar: $\leftrightarrow$}\ignorespaces}
\def\Mua{\marginpar[\boldmath\hfil$\Uparrow$]%
                   {\boldmath$\Uparrow$\hfil}%
                    \typeout{marginpar: $\uparrow$}\ignorespaces}
\def\Mda{\marginpar[\boldmath\hfil$\Downarrow$]%
                   {\boldmath$\Downarrow$\hfil}%
                    \typeout{marginpar: $\downarrow$}\ignorespaces}
\def\Mla{\marginpar[\boldmath\hfil$\Rightarrow$]%
                   {\boldmath$\Leftarrow $\hfil}%
                    \typeout{marginpar: $\leftrightarrow$}\ignorespaces}
\overfullrule 5pt
\oddsidemargin -15mm
\marginparwidth 29mm
}

\begin{document}

\thispagestyle{empty}
\def\thefootnote{\fnsymbol{footnote}}
\setcounter{footnote}{1}
\null
\strut\hfill FR-PHENO-2024-01, CERN-TH-2024-001
\vskip 0cm
\vfill
\begin{center}
{\large \bf 
\boldmath{Mixed NNLO QCD${}\times{}$electroweak corrections \\[.2em]
to single-Z production in pole approximation: \\[.2em]
differential distributions and forward--backward asymmetry
}
\par} \vskip 2.5em
{\large
{\sc Stefan Dittmaier$^1$, Alexander Huss$^2$
and Jan Schwarz$^1$}\\[1ex]
{\normalsize 
\it 
$^1$ Albert-Ludwigs-Universit\"at Freiburg, 
Physikalisches Institut, \\
Hermann-Herder-Stra\ss{}e 3,
D-79104 Freiburg, Germany \\[.5em]
$^2$ Theoretical Physics Department, CERN, 1211 Geneva 23, Switzerland
}
}

\par \vskip 1em
\end{center} \par
\vskip 2cm 
{\bf Abstract:} \par
Radiative corrections in pole approximation,
which are based on the leading contribution 
in a systematic expansion of amplitudes
about resonance poles, naturally decompose into \emph{factorizable} corrections
attributed to the production or decay of the resonance and 
\emph{non-factorizable} corrections induced by soft photon (or gluon) exchange
between those subprocesses.
In this paper we complete an earlier calculation of mixed QCD${}\times{}$electroweak
corrections of $\mathcal{O}(\alphas\alpha)$
to the neutral-current Drell--Yan cross section in pole
approximation by including the previously neglected corrections that are solely related to
the Z-boson production process.
We present numerical results both for differential distributions and for the
forward--backward asymmetry differential in the lepton-pair invariant mass,
which is the key observable in the measurement of the effective weak mixing angle
at the LHC.
Carefully disentangling the various types of factorizable and
non-factorizable corrections, we find 
(as expected in our earlier work)
that the by far most important contribution at $\mathcal{O}(\alphas\alpha)$
originates from the interplay of initial-state QCD corrections and
electroweak final-state corrections.
\par
\vfill
\noindent 
January 2024  \par
\vskip .5cm 
\null
\setcounter{page}{0}
\clearpage
\def\thefootnote{\arabic{footnote}}
\setcounter{footnote}{0}

\section{Introduction}
\label{se:intro}

Owing to its large cross section and clean experimental signature,
the Drell--Yan-like production of charged leptons is 
among the most important standard-candle processes at 
hadron colliders such as the Tevatron and the 
LHC~\cite{Dittmar:1997md,Khoze:2000db,Abdullin:2006aa,Gerber:2007xk}.
The charged-current channel (W~production) allows for
the determination of the mass and width of the 
W~boson~\cite{Haywood:1999qg,CDF:2013dpa,ATLAS:2017rzl,CDF:2022hxs,ATLAS:2023fsi},
the neutral-current channel (Z~production) for the measurement of the effective weak mixing 
angle~\cite{Haywood:1999qg,CMS:2011utm,ATLAS:2015ihy,LHCb:2015jyu,CDF:2018cnj,CMS:2018ktx}, 
both with extraordinary precision.
For these high-precision measurements, among the most relevant observables
are the transverse-momentum
and invariant-mass distributions, as well as the differential 
forward--backward asymmetry of the charged lepton pair from Z~production,
in the vicinity of the W- and Z-boson resonances.
Moreover, Drell--Yan cross sections significantly contribute
to the determination of the parton distribution functions (PDFs)
via rapidity distributions and the W-boson charge asymmetry~\cite{Boonekamp:2009yd}.
Last but not least, Drell--Yan-like processes are well suited to
search for new $\PW'$ and $\PZ'$~bosons in the high invariant-mass range
of the final-state leptons.

All these measurements and precision tests of the Standard Model require 
precise predictions for
differential Drell--Yan cross sections at the highest
possible level in order to match or better surpass the experimental uncertainties.
To this end, radiative corrections of the strong and electroweak (EW)
interactions as well as corrections mixing these types of interactions
have to be calculated to higher orders in perturbation theory.
Fixed-order QCD calculations are available fully differentially at
next-to-next-to-leading order (NNLO)~\cite{Hamberg:1990np,Gavin:2012sy,Gavin:2010az,Catani:2009sm,Melnikov:2006kv,Melnikov:2006di,Anastasiou:2003ds,Harlander:2002wh} with third-order results ($\textnormal{N}^3\textnormal{LO}$) known fully inclusively~\cite{Duhr:2020seh,Duhr:2021vwj,Baglio:2022wzu}, single-differentially~\cite{Chen:2021vtu,Chen:2022lwc}, and at the fiducial level~\cite{Chen:2022cgv,Neumann:2022lft,Chen:2022lwc}.
Additionally, the resummation of large QCD logarithms occurring due to
soft-gluon emissions at small transverse momentum has been studied in 
\citeres{Guzzi:2013aja,Kulesza:2001jc,Catani:2015vma,Balazs:1997xd,Landry:2002ix,%
Bozzi:2010xn,Mantry:2010mk,Becher:2011xn,Bizon:2018foh,Bizon:2019zgf,Re:2021con,Ju:2021lah,Camarda:2021ict,Neumann:2022lft},
and threshold effects have been calculated up to 
$\textnormal{N}^3\textnormal{LO}$ in QCD \cite{Ahmed:2014cla,Catani:2014uta}.
On the EW side, the next-to-leading order (NLO) corrections are
known~\cite{Baur:1997wa,Zykunov:2001mn,Baur:2001ze,Dittmaier:2001ay,Baur:2004ig,%
Arbuzov:2005dd,CarloniCalame:2006zq,Zykunov:2005tc,CarloniCalame:2007cd,Arbuzov:2007db,%
Brensing:2007qm,Dittmaier:2009cr,Barze:2013fru,Boughezal:2013cwa,Cieri:2018sfk,Autieri:2023xme}
as well as leading higher-order effects from multiple photon emissions or of universal 
origin~\cite{CarloniCalame:2007cd,Brensing:2007qm,Dittmaier:2009cr,Placzek:2003zg,%
CarloniCalame:2003ux}.

The mixing of QCD and EW corrections begins at NNLO, i.e.\ at $\mathcal{O}(\alphas\alpha)$.
The calculation of these corrections to W and Z~production
started in \citeres{Dittmaier:2014qza,Dittmaier:2015rxo}
in pole approximation (PA), which is based on the leading contribution in an expansion
of all matrix elements about the resonance poles
(see, e.g., \citere{Denner:2019vbn} for the general concept).
The PA reduces the complexity of the loop calculations, e.g.\
by avoiding two-loop box diagrams with the full $2\to2$ kinematics,
and classifies the $\mathcal{O}(\alphas\alpha)$ corrections into four separately 
gauge-invariant contributions~\cite{Dittmaier:2014qza}:
(i)~{\it factorizable initial--final (IF) corrections} including QCD corrections to W/Z production 
and EW corrections to the W/Z~decay, 
(ii)~{\it factorizable initial--initial (II) corrections} including mixed NNLO
QCD${}\times{}$EW corrections to on-shell W/Z~production,
(iii)~{\it final--final (FF) corrections} with QCD and EW corrections confined in the 
W/Z~decays, and 
(iv)~{\it non-factorizable (NF) corrections} linking the QCD-corrected W/Z~production to the 
leading-order (LO) decay process by soft photon exchange or emission.
The NF and FF corrections have been calculated in \citere{Dittmaier:2014qza}
and \citere{Dittmaier:2015rxo}, respectively, and found to be insignificant
in differential cross sections.
The corrections of type~IF have been evaluated in \citere{Dittmaier:2015rxo} as well
and found to be sizeable owing to the interplay of the large QCD corrections
in the production with enhanced photonic final-state radiation effects. 
These corrections, in particular, induce a shift in the W-boson mass extracted from
the charged-current process which was estimated to be about $10\MeV$.
The II~corrections had been neglected in \citeres{Dittmaier:2014qza,Dittmaier:2015rxo},
following arguments that they are expected to be subleading wrt.\ the IF~corrections
owing to the absorption of the enhanced collinear ISR effects into the PDFs in
contrast to the situation for FSR.
The missing II~corrections of $\mathcal{O}(\alphas\alpha)$ meanwhile have been
calculated in \citeres{Behring:2020cqi} and 
\cite{Bonciani:2016wya,deFlorian:2018wcj,Delto:2019ewv,Bonciani:2019nuy,%
Cieri:2020ikq,Buccioni:2020cfi,Bonciani:2020tvf,Bonciani:2021iis}
for W and Z~production, but unfortunately
have not yet been combined with the other correction types to a full prediction in PA.

\begin{sloppypar}
More recently, the $\mathcal{O}(\alphas\alpha)$ corrections to the full off-shell
Drell--Yan processes have been attacked by
several groups, starting with the $\mathcal{O}(N_f\alphas\alpha)$ corrections
to the charged- and neutral-current channels which are enhanced by the number $N_f$ of fermion 
flavours~\cite{Dittmaier:2020vra}.
For the off-shell charged-current process, the $\mathcal{O}(\alphas\alpha)$ corrections
have been evaluated in approximate form, taking into account real and real--virtual NNLO
corrections exactly, but neglecting the genuine two-loop 
corrections~\cite{Buonocore:2021rxx}.
For the off-shell neutral-current process, the full $\mathcal{O}(\alphas\alpha)$ corrections
have been calculated by two groups~\cite{Bonciani:2021zzf,Armadillo:2022bgm,Buccioni:2022kgy}.
While these achievements can certainly be considered as a major breakthrough in the
calculation of mixed QCD${}\times{}$EW corrections to $2\to2$ scattering processes,
we still see the need to complete the discussion of the phenomenological structure
and impact of the $\mathcal{O}(\alphas\alpha)$ corrections to Drell--Yan processes
at least in two respects.
Firstly, a thorough comparison of the approximate and full off-shell calculations with PA
predictions is very desirable, to better understand the origin of the dominant effects
and to obtain further guidance in the construction of approximations that are numerically
more efficient to evaluate.
Secondly, for the neutral-current case a proper phenomenological discussion of the
$\mathcal{O}(\alphas\alpha)$ corrections to the differential 
forward--backward asymmetry still does not exist in the literature.
\end{sloppypar}

In this paper we prepare for the first aspect by completing
the PA prediction started in \citeres{Dittmaier:2014qza,Dittmaier:2015rxo} by
calculating the missing II~corrections to Z~production.
Since \citere{Bonciani:2021zzf} employs the same setup as already used in
our previous calculation~\cite{Dittmaier:2014qza,Dittmaier:2015rxo},
we keep this setup in this paper as well and compare our PA results on
differential cross sections with the results from the off-shell calculation of
\citere{Bonciani:2021zzf}, which are based on bare muons.
A more complete comparison, including results based on dressed leptons
as used in \citere{Armadillo:2022bgm}, is beyond the scope of this paper
and will certainly carried out within the LHC Electroweak Working Group in the
near future.
The major part of our discussion of numerical results is devoted to the 
$\mathcal{O}(\alphas\alpha)$ corrections to the forward--backward asymmetry
in the Z~resonance region in the neutral-current Drell--Yan process.
This discussion is of particular relevance for the theory predictions
to the experimental determination of the effective weak mixing angle at the LHC.

This paper is organized as follows: 
In \refse{se:details-calc} we give a short overview of the gauge-invariant PA
contributions to the $\mathcal{O}(\alphas\alpha)$ corrections and
discuss our calculation of contributions of type~II.
The latter are divided into mixed QCD${}\times{}$weak and QCD${}\times{}$photonic corrections in a
gauge-invariant way on the basis of selecting appropriate subsets of diagrams.
The QCD${}\times{}$weak part comprises only genuine two-loop corrections to the
$\PZ\bar ff$ vertex and real--virtual corrections with jet emission.
Our result for the corresponding two-loop formfactor, which was
first calculated in \citere{Kotikov:2007vr}, 
is presented in \refapp{se:app:formfactors}.
Since the infrared (IR) singularities in the QCD${}\times{}$weak corrections are only of
NLO complexity, we apply both the antenna~\cite{Daleo:2006xa} and
dipole subtraction~\cite{Catani:1996vz} approaches
to combine the two IR-singular parts to a total IR-finite result.
The QCD${}\times{}$photonic corrections feature double-virtual,
real--virtual, and double-real corrections. 
To isolate and combine all NNLO IR singularities, which is the major complication
in this part, we apply antenna subtraction as introduced in \citeres{Gehrmann-DeRidder:2005btv,Currie:2013vh}.
Our calculation is the first application of antenna subtraction to
$\mathcal{O}(\alphas\alpha)$ corrections. 
In \refse{se:num-res} we present a detailed discussion of the 
$\mathcal{O}(\alphas\alpha)$ corrections to neutral-current DY processes in the 
Z-resonance region on various distributions, with special emphasis on the newly 
calculated corrections to the forward--backward asymmetry. 
To complete the phenomenological picture we also provide full NLO results
(i.e.\ without resorting to the PA), split into a genuine weak part
and photonic parts induced by initial-state radiation, final-state radiation,
and initial--final interference.
The PA-based $\mathcal{O}(\alphas\alpha)$ correction is split into
II, FF, IF, and NF parts as introduced above.
Finally, we also evaluate the relevant leading EW effects beyond NNLO, which
are induced by multi-photon radiation and the leading universal EW renormalization effects.
Our summary is given in \refse{se:summary}, and the appendices provide
further analytical results.

\section{Details of the calculation}
\label{se:details-calc}

In this section we describe the calculation of $\mathcal{O}(\alphas\alpha)$ corrections to the neutral-current Drell--Yan process in \PPA. In \refse{se:survey-diags} we give an overview of the separately gauge-invariant building blocks of the \PPA\ at $\O(\alphas\alpha)$. 
Corrections of initial--initial type---i.e.\ the $\mathcal{O}(\alphas\alpha)$ corrections 
with both QCD and EW corrections to the production of the $\PZ$~boson---are 
the last missing piece to complete the calculation~\cite{Dittmaier:2014qza,Dittmaier:2015rxo} 
of $\mathcal{O}(\alphas\alpha)$ corrections to DY-like $\PZ$-boson production in \PPA;
their calculation is described in \refse{se:calc-ii}.

\subsection{{\boldmath{$\mathcal{O}(\alphas \alpha)$}} corrections to single-{$\PZ$} production in pole approximation}
\label{se:survey-diags}

\begin{figure}
  \centering
  \begin{subfigure}[m]{.48\linewidth}
    \centering
    \includegraphics{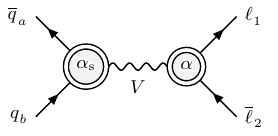}
    \subcaption{Factorizable initial--final (IF) corrections}
  \end{subfigure}
\hspace{.5em}
  \begin{subfigure}[m]{.48\linewidth}
    \centering
    \includegraphics{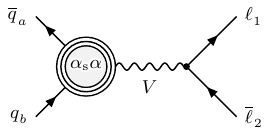}
    \subcaption{Factorizable initial--initial (II) corrections}
  \end{subfigure}
  \\[1.5em]
  \begin{subfigure}[m]{.48\linewidth}
    \centering
    \includegraphics{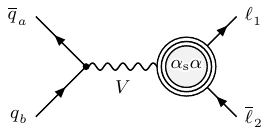}
    \subcaption{Factorizable final--final (FF) corrections}
  \end{subfigure}
\hspace{.5em}
  \begin{subfigure}[m]{.48\linewidth}
    \centering
    \includegraphics{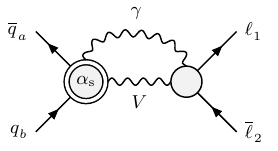}
    \subcaption {Non-factorizable (NF) corrections}
  \end{subfigure}
  \caption{The four types of corrections that contribute to the mixed QCD${}\times{}$EW corrections in PA illustrated in terms of generic two-loop amplitudes. Simple circles symbolize tree structures, double circles one-loop corrections, and triple circles two-loop contributions.}
  \label{fig:NNLOcontrib}
\end{figure}

The expansion of the full NNLO $\mathcal{O}(\alphas \alpha)$ correction around the $\PZ$  resonance pole at $p_Z^2 \approx M_Z^2$ leads to the following four types of corrections \cite{Dittmaier:2014qza,Dittmaier:2015rxo}, which are illustrated for the purely virtual two-loop corrections
in \reffi{fig:NNLOcontrib}:
\begin{itemize}
\item 
Factorizable $\mathcal{O}(\alphas \alpha)$ corrections of initial--final (IF) type
combine the $\mathcal{O}(\alphas)$ corrections to $\PZ$~production and 
the $\mathcal{O}(\alpha)$ correction to the leptonic $\PZ$~decay. 
Here and in the following the terminology
``factorizable'' refers to the fact that the corresponding amplitudes all
factorize in terms of subamplitudes for production and decay and a resonant $\PZ$-boson
propagator.
The strong collinear enhancement of final-state photon radiation renders 
the IF~class of $\mathcal{O}(\alphas \alpha)$ corrections by far the dominant 
PA contribution in the resonance region~\cite{Dittmaier:2015rxo}. 
 \item  
The factorizable initial--initial (II) $\mathcal{O}(\alphas \alpha)$ corrections
contain contributions where both QCD and EW corrections are located in the $\PZ$-boson 
production subprocess. This contribution is essentially furnished by the corrections
to on-shell $\PZ$~production, supplemented by the off-shell $\PZ$~propagator and the
leptonic $\PZ$~decay in LO.
Since both QCD and photonic effects from initial-state radiation are widely
absorbed into PDFs, the type-II $\mathcal{O}(\alphas \alpha)$ corrections were
expected to be suppressed wrt.\ to the dominating IF corrections and neglected
in \citeres{Dittmaier:2014qza,Dittmaier:2015rxo}. In this paper, we complete
the PA at $\mathcal{O}(\alphas \alpha)$ by supplementing the corrections of type~II.
We split the II~corrections 
into the separately gauge-invariant QCD$\times$weak and QCD$\times$photonic parts
of the orders $\mathcal{O}(\alphas \alpha_\Pw)$ and
$\mathcal{O}(\alphas\alpha_\text{p})$, respectively.
The photonic initial-state corrections are identified
as the part of the $\mathcal{O}(\alpha)$ corrections that are proportional to
the product of quark charges and comprise all contributions where
the photon couples only to the quark or antiquark.
Note that the QCD$\times$photonic corrections even form
a gauge-invariant part of the full off-shell calculation without PA.
\item 
The factorizable final--final (FF) $\mathcal{O}(\alphas \alpha)$ corrections include only 
counterterm corrections to the lepton--$\PZ$~vertices
and, in particular, do not receive contributions from real gluon or photon radiation.
In \cite{Dittmaier:2015rxo} the explicit calculation of these corrections 
confirmed the expectation that their impact on differential cross sections
is phenomenologically negligible.
\item 
Non-factorizable (NF) corrections are induced by soft-photon exchange or emission
connecting final-state leptons and initial-state quarks, combined with QCD 
$\mathcal{O}(\alphas)$ corrections to $\PZ$-boson production. 
The non-trivial momentum flow of the soft photon between production and decay subprocesses,
which implies that the squared matrix elements are not proportional to a
squared $\PZ$~propagator, gave rise to the terminology ``non-factorizable''.
Owing to a systematic cancellation between real and virtual NF
corrections the numerical impact of these corrections on differential cross sections
is at the sub-permille 
level~\cite{Dittmaier:2014qza} and therefore of no relevance for phenomenology.
\end{itemize}
The four types of $\O(\alphas\alpha)$ corrections in PA can be 
further classified into the usual NNLO contributions of types
double-real, real--virtual, and double-virtual.
Figure~\ref{fig:NNLOcontrib} shows the separation of the double-virtual corrections
into IF, II, FF, and NF contributions; the corresponding separation of the
double-real and real--virtual corrections is obvious.

\subsection{Calculation of the factorizable initial--initial corrections}
\label{se:calc-ii}

In this section we present the details of our calculation of the corrections of type~II.
The calculation of the corrections of types~IF+FF and type~NF can be found in
\citeres{Dittmaier:2015rxo} and \cite{Dittmaier:2014qza}, respectively.
Since the PA for the factorizable corrections is based on amplitudes for the
production and decay subprocesses, the implementation of the PA involves
a projection of momenta to on-shell (OS) $\PZ$~bosons in the subamplitudes 
(but not in the intermediate $\PZ$~propagator), in order to maintain gauge invariance in the
subamplitudes.
The details of the OS mappings are discussed when they become relevant below.

The $\text{QCD}\times\text{photonic}$ 
$\mathcal{O}(\alphas \alpha_\text{p})$ corrections of type~II 
are proportional 
to the charge factors $Q_q^2$ of the (anti)quarks
and therefore gauge invariant even without applying the \PPA. 
In order to stay closer to the full calculation, we evaluate the $\text{QCD}\times\text{photonic}$
corrections without employing the PA.
However, for the $\text{QCD}\times\text{weak}$ $\mathcal{O}(\alphas \alpha_\Pw)$ 
corrections we have to use the \PPA\ to preserve gauge invariance.
Both for the $\text{QCD}\times\text{weak}$ 
and the $\text{QCD}\times\text{photonic}$ II~corrections 
we have performed two independent calculations which produce results that
are in mutual numerical agreement.

\begin{figure}
\begin{subfigure}{1.0\textwidth}
\centering
\mbox{\includegraphics{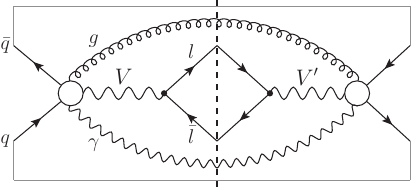}
\qquad
\includegraphics{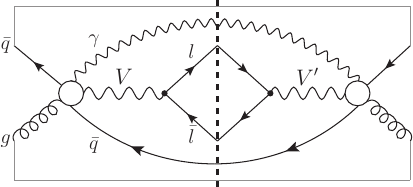}
}
\\[.5em]
\centering
\includegraphics{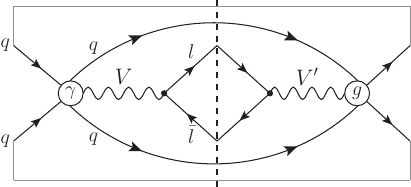}
\caption{Double-real $\mathcal{O}(\alphas\alpha_\text{p})$ II interference diagrams}
\label{fig:rr-diags}
\end{subfigure}
\\[1em]
\begin{subfigure}{1.0\textwidth}
\centering
\mbox{\includegraphics{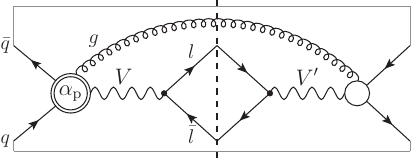}
\qquad
\includegraphics{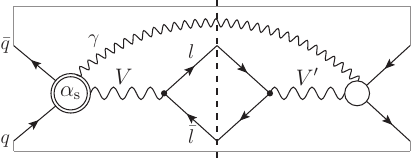}
}
\\[.5em]
\centering
\includegraphics{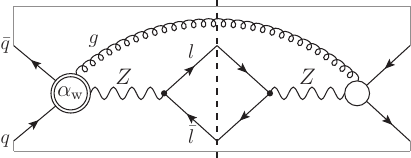}
\caption{Real--virtual $\mathcal{O}(\alphas\alpha_\text{p})$ II (first line) and $\cal O(\alphas\alpha_\text{w})$ II (second line) interference diagrams}
\label{fig:rv-diags}
\end{subfigure}
\\[1em]
\begin{subfigure}{1.0\textwidth}
\centering
\includegraphics{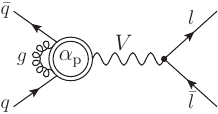}
\qquad
\includegraphics{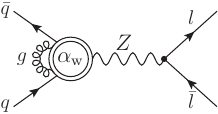}
% \qquad
% \includegraphics{images/diags/DY_red3}
\caption{Double-virtual $\mathcal{O}(\alphas\alpha_\text{p})$ and $\cal O(\alphas\alpha_\text{w})$ 
II diagrams}
\label{fig:vv-a-diags}
\end{subfigure}
\caption{Various contributions to the gauge-invariant set of 
$\mathcal{O} (\alphas\alpha_\text{p})$ and $\cal O(\alphas\alpha_\text{w})$ II~corrections, 
where $V,V^\prime=Z,\gamma$. Double circles indicate one-loop corrections, simple circles 
indicate relevant tree structures, and simple circles with a ``$\gamma$'' (``$g$'') inside 
represent all possible connected tree-level diagrams of the process 
$q_a q_a \to q_a q_a + V$ with an intermediate photon (gluon). 
An additional particle attached to a ``one-loop blob'', as e.g.\ in \reffi{fig:vv-a-diags}, 
means that the particle has to be inserted into the corresponding one-loop diagram 
in all possible ways.}
\label{fig:ii-diags}
\end{figure}

The double-real, real--virtual, and double-virtual contributions arising in the 
calculation of corrections of type~II are depicted in \reffi{fig:ii-diags}. 
Initial--initial $\mathcal{O} (\alphas\alpha_\text{p})$ corrections demand a proper NNLO
subtraction scheme as they involve two potentially unresolved particles in the final state.
To this end, we employ antenna subtraction. 
The construction of antenna subtraction functions at $\mathcal{O} (\alphas\alpha_\text{p})$ 
is based on the subleading colour parts of the known $\mathcal{O} (\alphas^2)$ antenna 
functions for the initial--final~\cite{Daleo:2009yj} and 
initial--initial~\cite{Gehrmann:2011wi,Gehrmann-DeRidder:2012too} cases.
In contrast to the $\text{QCD}\times\text{photonic}$ II~corrections, 
in the $\text{QCD}\times\text{weak}$ II~corrections
only one potentially unresolved particle is involved, 
and therefore the IR pole structure is of one-loop complexity, 
so that one-loop subtraction schemes are sufficient to handle the IR poles.
In this case we have applied both antenna~\cite{Daleo:2006xa}
and dipole subtraction~\cite{Catani:1996vz}
and compared the obtained results, which are in agreement.

\subsubsection{Double-virtual corrections}
\label{se:calc-ii-vv}

The double-virtual corrections affect the underlying $2\to2$ process
\begin{equation}
  q(p_\Pq,\sigma_\Pq) + \bar q(p_{\bar\Pq},\sigma_{\bar\Pq}) \;\to\; 
\Pl^-(k_\ell,\tau_\ell) + \Pl^+(k_{\bar\ell},\tau_{\bar\ell}), 
\end{equation}
where the momenta and helicity labels of the respective particles are given in
parentheses. The total incoming momentum is
denoted $q=p_\Pq+p_{\bar\Pq}$ in the following.
All external fermions are taken massless, i.e.\
$p_\Pq^2=p_{\bar\Pq}^2=k_\ell^2=k_{\bar\ell}^2=0$. In the following we will make use of
the Mandelstam variables
\begin{equation}
  \hat s = (p_\Pq+p_{\bar\Pq})^2 = q^2, \quad
  \hat t = (p_\Pq-k_\ell)^2, \quad
  \hat u = (p_\Pq-k_{\bar\ell})^2.
\end{equation}

The double-virtual II~corrections to the $\Pq\bar\Pq\rightarrow \Pl^- \Pl^+$
matrix element receive contributions from $\text{QCD}\times\text{weak}$
$\mathcal{O}(\alpha_\Ps\alpha_\Pw)$ and $\text{QCD}\times\text{photonic}$
$\mathcal{O}(\alpha_\Ps\alpha_\text{p})$ corrections. 
The $\mathcal{O}(\alpha_\Ps\alpha_\Pw)$ corrections to the $\PZ\bar\Pq\Pq$ vertex 
are contained in the two-loop contribution to the 
renormalized $\PZ\bar\Pq\Pq$ formfactors
$\hat F^{Z\bar\Pq\Pq}_{\pm}(q^2)$ for light quarks,
\begin{equation}
\Gamma^{Z\bar\Pq\Pq}_{\rR,\mu}(-q,p_{\bar q},p_q) = 
e \sum_{\tau=\pm} \hat F^{Z\bar\Pq\Pq}_{\tau}(q^2) \gamma_\mu \omega_\tau, \qquad
\omega_\pm = \frac{1}{2}(1\pm\gamma_5),
\label{eq:Faas}
\end{equation}
where $\Gamma^{Z\bar\Pq\Pq}_{\rR,\mu}$ is the renormalized $\PZ\bar\Pq\Pq$ vertex function with
off-shell $\PZ$~momentum $q$ projected onto massless external on-shell 
(anti)quarks~$\Pq/\bar\Pq$ and $e$ is the elementary charge.
For field-theoretical quantities and Standard Model fields and parameters we follow
the notation and conventions of \citere{Denner:2019vbn}, if not defined otherwise.
The LO contributions $F^{Z\bar\Pq\Pq}_{\LO,\tau}$ to the formfactors are given by
\begin{align}
F^{Z\bar\Pq\Pq}_{\LO,\tau} = g_q^\tau, 
\end{align}
with the chiral couplings
 \begin{equation}
  g_f^+ = -\frac{Q_f \sw}{\cw}, \qquad 
g_f^-= \frac{I^3_{\rw,f}- Q_f s_\rw^2}{\cw \sw},
 \end{equation}
where $Q_f$ and $I^3_{\rw,f}$ denote the electric charge and the third component of weak
isospin of some fermion~$f$, respectively.
Here, $\sw$ and $\cw$ are the sine and cosine of the weak mixing angle, which
is related to the masses $\MW$ and $\MZ$ of the W/Z bosons according to
$\cw=\MW/\MZ$.

It is convenient to define the so-called reducible (red) parts
of the $\mathcal{O}(\alpha_\Ps\alpha_\Pw)$ contribution to the formfactors 
as the products of the renormalized one-loop QCD and weak contributions to the formfactors,
\begin{equation}
\hat F^{Z\bar\Pq\Pq,\text{red}}_{\text{V}_\Ps\otimes\text{V}_\Pw,\tau}(q^2) = 
\delta_{\text{V}_\Ps}^{Z\bar\Pq\Pq}(q^2) \,  
\hat F^{Z\bar\Pq\Pq}_{\text{V}_\Pw,\tau}(q^2),
\end{equation}
where $\delta_{\text{V}_\Ps}^{Z\bar\Pq\Pq}(q^2)$ is the renormalized NLO
QCD correction factor, e.g., given in Eq.~(2.35) of 
\citere{Dittmaier:2014qza},
and the renormalized NLO weak formfactor contribution
$\hat F^{Z\bar\Pq\Pq}_{\text{V}_\Pw,\tau}(q^2)$.
The latter decomposes into unrenormalized part 
$F^{Z\bar\Pq\Pq}_{\text{V}_\Pw,\tau}(q^2)$ and counterterm
contribution $\delta^{\text{ct},\tau}_{Z\bar qq,\text{weak}}$,
\begin{equation}
\hat F^{Z\bar\Pq\Pq}_{\text{V}_\Pw,\tau}(q^2) =
F^{Z\bar\Pq\Pq}_{\text{V}_\Pw,\tau}(q^2)+ 
F^{Z\bar\Pq\Pq}_{\LO,\tau}\, \delta^{\text{ct},\tau}_{Z\bar qq,\text{weak}}.
\end{equation}
Explicit expressions for the weak corrections and counterterms can, e.g.,  
be found in \citeres{Dittmaier:2009cr}.

\begin{figure}
\begin{subfigure}{1.0\textwidth}
\centering
\mbox{\includegraphics[scale=1.25]{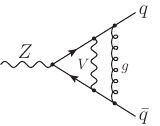}
\qquad
\includegraphics[scale=1.25]{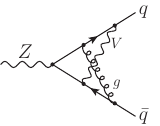}
}
\\[.5em]
\centering
\mbox{\includegraphics[scale=1.25]{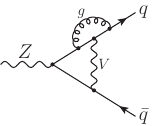}
\qquad
\includegraphics[scale=1.25]{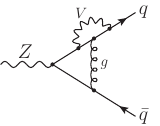}
}
\\[.5em]
\centering
\mbox{\includegraphics[scale=1.25]{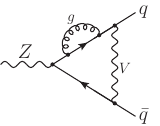}
\qquad
\includegraphics[scale=1.25]{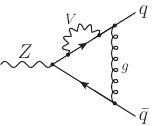}
\qquad
\includegraphics[scale=1.25]{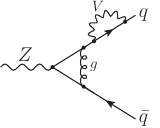}
% \qquad
% \includegraphics{images/Vself_Gvert_3}
}
\caption{Abelian diagrams}
\label{fig:irred-diags-abelian}
\end{subfigure}
\\[1em]
\begin{subfigure}{1.0\textwidth}
\centering
\mbox{\includegraphics[scale=1.25]{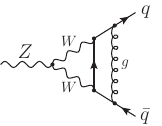}
\qquad
\includegraphics[scale=1.25]{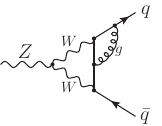}
\qquad
\includegraphics[scale=1.25]{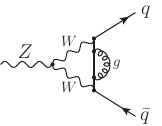}
}
\caption{Non-Abelian diagrams}
\label{fig:irred-diags-non-abelian}
\end{subfigure}
\caption{Example diagrams for the various $\cal O(\alphas\alpha_\text{w})$ two-loop contributions to the $\PZ\bar\Pq\Pq$ vertex, where $V=Z,W$.}
\label{fig:irred-ii-diags}
\end{figure}%
To extract the genuine NNLO information contained in the full 
$\mathcal{O}(\alpha_\Ps\alpha_\Pw)$
formfactor we define the irreducible (irred) contribution as the 
difference between the full $\mathcal{O}(\alpha_\Ps\alpha_\Pw)$ 
and the reducible contributions to the formfactor,
\begin{equation}
\hat F^{Z\bar\Pq\Pq,\text{irred}}_{\text{V}_\Ps\otimes\text{V}_\Pw,\tau}(q^2)
= \hat F^{Z\bar\Pq\Pq}_{\text{V}_\Ps\otimes\text{V}_\Pw,\tau}(q^2)
-\hat F^{Z\bar\Pq\Pq,\text{red}}_{\text{V}_\Ps\otimes\text{V}_\Pw,\tau}(q^2),
\end{equation}
where the corresponding diagrams are shown in \reffi{fig:irred-ii-diags}.
This irreducible contribution further decomposes into a 
($q^2$-independent) counterterm part 
$\hat F^{Z\bar\Pq\Pq,\text{irred}}_{N_f\alphas\alpha_\Pw,\tau}$
containing all irreducible two-loop contributions with closed quark loops 
and an internal gluon and a part
$F^{Z\bar\Pq\Pq,\text{irred}}_{\text{V}_\Ps\otimes\text{V}_\Pw,\tau}(q^2)$
comprising all genuine two-loop corrections to the vertex and the external
quark lines.
For the $N_f$-enhanced counterterm part 
$\hat F^{Z\bar\Pq\Pq,\text{irred}}_{N_f\alphas\alpha_\Pw,\tau}$
we adopt the results of
\citere{Dittmaier:2020vra}, where the $\mathcal{O}(N_f\alpha_\Ps\alpha)$ corrections
to the full off-shell process have been calculated,
\begin{equation}
\hat F^{Z\bar\Pq\Pq,\text{irred}}_{N_f\alphas\alpha_\Pw,\tau}
= F^{Z\bar\Pq\Pq}_{\LO,\tau}\,\delta^{\mathrm{ct},\tau}_{Z\bar ff,(\alphas\alpha)},
\end{equation}
where the counterterm correction factor 
$\delta^{\mathrm{ct},\tau}_{Z\bar ff,(\alphas\alpha)}$ is given in Eq.~(2.46) of
\citere{Dittmaier:2020vra}.%
\footnote{Actually, the result given there is formulated in the complex-mass scheme.
The translation to the real OS renormalization scheme used in this paper
is, however, obvious.}
The $N_f$-independent $\mathcal{O}(\alphas\alpha_\Pw)$ part 
$F^{Z\bar\Pq\Pq,\text{irred}}_{\text{V}_\Ps\otimes\text{V}_\Pw,\tau}(q^2)$
is most conveniently formulated with 
two auxiliary functions, called $\phi_\PA$ and $\phi_\text{NA}$ in the following,
which have been introduced and calculated in \citere{Kotikov:2007vr}, 
 \begin{equation}
 \begin{aligned}
  F^{Z\bar\Pq\Pq,\text{irred}}_{\text{V}_\Ps\otimes\text{V}_\Pw,+}(q^2) ={}  & 
\CF \frac{\alpha_s}{4 \pi}
  \frac{\alpha}{4 \pi} (g_q^+)^3\,\phi_\PA(q^2/\MZ^2), \\
  F^{Z\bar\Pq\Pq,\text{irred}}_{\text{V}_\Ps\otimes\text{V}_\Pw,-}(q^2) ={} & 
\CF \frac{\alpha_s}{4 \pi}
  \frac{\alpha}{4 \pi} \left(  (g_q^-)^3 \phi_\PA(q^2/\MZ^2) + \frac{g_q^-}{2  \sw^2} \phi_\PA(q^2/\MW^2)
  + I^3_{\rw,\Pq} \frac{\cw}{2  \sw^3} \phi_\text{NA}(q^2/\MW^2) \right),
 \end{aligned}
\label{eq:FZff}
 \end{equation}
where $\CF=4/3$.

\begin{sloppypar}
We have performed a completely independent recalculation of 
the formfactors $\phi_\PA$ and $\phi_\text{NA}$.
The graphs for the two-loop vertex corrections 
were generated with 
{\sc FeynArts}$1.0$~\cite{Kublbeck:1990xc}. To express the amplitudes in terms of scalar two-loop master integrals the amplitudes were further algebraically reduced with inhouse {\sc Mathematica} routines 
combined with
an integral reduction based on 
\KIRA~\cite{Maierhoefer:2017hyi,Maierhofer:2018gpa}. The scalar master integrals were calculated via differential equations~\cite{Henn:2013pwa,Henn:2014qga} producing results in terms of Goncharov Polylogarithms (GPLs) \cite{Goncharov:1998kja,Goncharov:2010jf}.
The GPLs are evaluated with an inhouse library that was checked against results from
the programs {\sc CHAPLIN}~\cite{Buehler:2011ev} and {\sc handyG}~\cite{Naterop:2019xaf}.
Further details of our formfactor calculation as well as benchmark numbers
are given in \refapp{se:app:formfactors}.
Our second, independent calculation of the $\O(\alphas\alpha_\Pw)$ corrections
makes use of the results for the unrenormalized irreducible contribution to the
formfactor as given in \citere{Kotikov:2007vr}.
Appendix~\ref{se:app:formfactors} also reports on a comparison between our
formfactor results and the ones taken from \citere{Kotikov:2007vr},
revealing numerical agreement if we numerically evaluate the analytical
results of \citere{Kotikov:2007vr}, although we cannot fully reproduce all
benchmark numbers given there.
\end{sloppypar}

In summary, the 
$\mathcal{O}(\alpha_\Ps\alpha_\Pw)$ contributions to the $\PZ\bar\Pq\Pq$ formfactor
defined in \refeq{eq:Faas}
decompose as 
\begin{align}
\hat F^{Z\bar\Pq\Pq}_{\alphas\alpha_\Pw,\tau}(q^2) = {}&
\hat F^{Z\bar\Pq\Pq,\text{red}}_{\text{V}_\Ps\otimes\text{V}_\Pw,\tau}(q^2)
+ \hat F^{Z\bar\Pq\Pq,\text{irred}}_{\text{V}_\Ps\otimes\text{V}_\Pw,\tau}(q^2)
\nn\\
= {}&
\delta_{\text{V}_\Ps}^{Z\bar\Pq\Pq}(q^2) 
\left[F^{Z\bar\Pq\Pq}_{\text{V}_\Pw,\tau}(q^2)+ 
F^{Z\bar\Pq\Pq}_{\LO,\tau}\, \delta^{\text{ct},\tau}_{Z\bar qq,\text{weak}}\right]
+ F^{Z\bar\Pq\Pq}_{\LO,\tau}\,\delta^{\mathrm{ct},\tau}_{Z\bar ff,(\alphas\alpha)}
+ F^{Z\bar\Pq\Pq,\text{irred}}_{\text{V}_\Ps\otimes\text{V}_\Pw,\tau}(q^2).
\end{align}
In all contributions from closed quark loops, the full dependence on the bottom-
and top-quark masses is kept. All other appearances of bottom quarks are connected
to external quarks, which are all taken massless. In the one-loop formfactor
$F^{Z\bar\Pq\Pq}_{\text{V}_\Pw,\tau}(q^2)$ and in the irreducible two-loop
contribution $F^{Z\bar\Pq\Pq,\text{irred}}_{\text{V}_\Ps\otimes\text{V}_\Pw,\tau}(q^2)$,
W-boson exchange leads to the appearance of top-quarks in the $\Pb\bar\Pb$~channel,
which is suppressed wrt.\ the other $q\bar q$ channels.
In $F^{Z\bar\Pq\Pq}_{\text{V}_\Pw,\tau}(q^2)$ the corresponding dependence on the top-quark
mass $\Mt$ is kept, while $\Mt$ is set to zero in 
$F^{Z\bar\Pq\Pq,\text{irred}}_{\text{V}_\Ps\otimes\text{V}_\Pw,\tau}(q^2)$.
To assess the validity of this approximation, we have also set $\Mt$ to zero
in $F^{Z\bar\Pq\Pq}_{\text{V}_\Pw,\tau}(q^2)$ as well, which changes the 
$\text{QCD}\times\text{weak}$ corrections to the $M_{\ell\ell}$ distribution
by less than $0.05\%$ and to the 
forward--backward asymmetry $A_\text{FB}(M_{\ell\ell})$ by about $10^{-5}$ at most,
which is both phenomenologically completely negligible.

The $\mathcal{O}(\alphas\alpha_\Pw)$ correction of type~II
to the squared $\Pq\bar\Pq\rightarrow \ell\bar \ell$ amplitude, 
$M^{\Pq\bar\Pq\rightarrow \ell\bar \ell}_{\text V_\Ps \otimes \text V _\Pw,\text{II},\PPA}$,
is obtained from the interference between the genuine two-loop 
$\mathcal{O}(\alphas\alpha_\Pw)$ matrix element 
$\M^{\Pq\bar\Pq\rightarrow \ell\bar \ell}_{\text V_\Ps \otimes \text V_\Pw,\text{II},\PPA}$
and the LO matrix element $\M_{\LO,Z}^{\Pq\bar\Pq\rightarrow \ell\bar \ell}$ and 
the interference of two one-loop matrix elements
$\M^{\Pq\bar\Pq\rightarrow \ell\bar \ell}_{\text V_\Ps,\text{I},\PPA}$ and
$\M_{\text V_\Pw,\text{I},\PPA}^{\Pq\bar\Pq\rightarrow \ell\bar \ell}$, 
the former with an $\mathcal{O}(\alphas)$ and the latter with an
$\mathcal{O}(\alpha_\Pw)$ initial-state correction, 
\begin{equation}
 \label{eq:II:NNLO-vv-asaw}
 M^{\Pq\bar\Pq\rightarrow \ell\bar \ell}_{\text V_\Ps \otimes \text V _\Pw,\text{II},\PPA}
  = 2 \Re \Big\{ \M^{\Pq\bar\Pq\rightarrow \ell\bar \ell}_{\text V_\Ps \otimes \text V_\Pw,\text{II},\PPA} 
\left( \M_{\LO,Z}^{\Pq\bar\Pq\rightarrow \ell\bar \ell} \right)^* 
+ \M^{\Pq\bar\Pq\rightarrow \ell\bar \ell}_{\text V_\Ps,\text{I},\PPA} 
\left( \M_{\text V_\Pw,\text{I},\PPA}^{\Pq\bar\Pq\rightarrow \ell\bar \ell} \right)^* \Big\}.
\end{equation}
In PA, the individual matrix elements in the last equation are obtained by employing the 
respective on-shell form factors $F_{\dots}^{Z\bar qq}(q^2=M_Z^2)$, 
\begin{align}
\M_{\LO,Z}^{\Pq\bar\Pq\rightarrow \ell\bar \ell} ={}& e^2\, 
\frac{F^{Z\bar\Pq\Pq}_{\LO,\sigma}\, g_\ell^\tau}{\hat s-\mu_Z^2}
\,\mathcal{A}^{\sigma\tau},
\\
\M_{\text V_\Pw,\text{I},\PPA}^{\Pq\bar\Pq\rightarrow \ell\bar \ell} ={}& e^2\, 
\frac{\hat F^{Z\bar\Pq\Pq}_{\text{V}_\Pw,\sigma}(\MZ^2) \,
g_\ell^\tau}{\hat s-\mu_Z^2}
\,\mathcal{A}^{\sigma\tau}, \qquad
\M_{\text V_\Ps,\text{I},\PPA}^{\Pq\bar\Pq\rightarrow \ell\bar \ell} =
\delta_{\text{V}_\Ps}^{Z\bar\Pq\Pq}(\MZ^2) \,
\M_{\LO,Z}^{\Pq\bar\Pq\rightarrow \ell\bar \ell},
\\
\M^{\Pq\bar\Pq\rightarrow \ell\bar \ell}_{\text V_\Ps \otimes \text V_\Pw,\text{II},\PPA} 
={}& e^2\,
\frac{\hat F^{Z\bar\Pq\Pq}_{\alphas\alpha_\Pw,\sigma}(\MZ^2)\, g_\ell^\tau}{\hat s-\mu_Z^2}
\,\mathcal{A}^{\sigma\tau},
\label{eq:ampli-virt-ZPA-PA}
\end{align}
ensuring gauge invariance of the respective corrections. 
In the last equation $\mu_Z^2=\MZ^2-\ri\MZ\Gamma_\PZ$ is the gauge-independent 
location of the \PZ-propagator pole, and $\mathcal{A}^{\sigma\tau}$ are Dirac chains
containing the information on the chiralities of the quark and lepton spinor
chains $\sigma$ and $\tau$, respectively; in \citere{Dittmaier:2009cr} the
$\mathcal{A}^{\sigma\tau}$ were calculated to
\begin{align}
  \mathcal{A}^{\pm\pm}=2 \hat u, \qquad \mathcal{A}^{\pm\mp}=2 \hat t.
  \label{eq:dirac-struc-virt-ZPA}
\end{align}
Recall that $\sigma$ and $\tau$ fix the helicities of the external fermions
($\sigma=\sigma_\Pq=-\sigma_{\bar\Pq}, \tau=\tau_\ell=-\tau_{\bar\ell}$), which
are taken massless, so that the matrix-element contributions for different
combinations $\sigma\tau$ do not interfere.

Equation~\refeq{eq:II:NNLO-vv-asaw} contains
products of weak and QCD one-loop corrections, raising the question whether
the one-loop corrections are needed to higher orders in $\epsilon=(4-D)/2$
in dimensional regularization.
This is, however, not the case, i.e.\ it is sufficient to evaluate
all one-loop corrections to ${\cal O}(\epsilon^0)$.
Since the weak one-loop corrections are finite, no ${\cal O}(\epsilon)$
terms of the QCD one-loop corrections can produce relevant terms.
On the other hand, the weak one-loop corrections eventually multiply
the finite sum of QCD one-loop corrections and the integrated contributions
of the QCD subtraction function (from antenna or dipole subtraction),
so that ${\cal O}(\epsilon)$ terms of the weak one-loop corrections cannot 
produce relevant terms either.

The contribution of double-virtual $\QCD \times \text{photonic}$ 
corrections to the squared $\Pq\bar\Pq\rightarrow \ell\bar \ell$ amplitude is given by the
interference of the genuine two-loop $\mathcal{O}(\alpha_\Ps\alpha_\text{p})$ 
II~amplitude and the $\LO$ amplitude, and by the 
interference between two one-loop amplitudes with initial-state corrections,
\begin{equation}
 \label{eq:II:NNLO-vv-asap}
  M^{\Pq\bar\Pq\rightarrow \ell\bar \ell}_{\text V_\Ps \otimes \text V _\text{p},\text{II}}
  =  2 \Re \Big\{ \M^{\Pq\bar\Pq\rightarrow \ell\bar \ell}_{\text V_\Ps \otimes \text V_\text{p},\text{II}} \left( \M_{\LO,Z/\gamma}^{\Pq\bar\Pq\rightarrow \ell\bar \ell} \right)^* 
  + \M^{\Pq\bar\Pq\rightarrow \ell\bar \ell}_{\text V_\Ps,\text{I}} \left( \M_{\text V_\text{p},\text{I}}^{\Pq\bar\Pq\rightarrow \ell\bar \ell} \right)^* \Big\}.
\end{equation}
Owing to parity invariance of QCD and QED, the QCD, the photonic, and the 
$\QCD \times \text{photonic}$ correction factors to the right- and left-handed
formfactors of the $\PZ\bar\Pq\Pq$ vertex coincide, so that the corrected matrix
elements are proportional to the LO amplitude,
\begin{align}
2 \Re \Big\{
\M^{\Pq\bar\Pq\rightarrow \ell\bar \ell}_{\text V_\Ps \otimes \text V_\text{p},\text{II}} 
\left( \M_{\LO,Z/\gamma}^{\Pq\bar\Pq\rightarrow \ell\bar \ell} \right)^*
\Big\} &= 
\delta_{\text{V}_\Ps\otimes\text{V}_\text{p}}^{Z\bar\Pq\Pq,[2 \times 0]}(\hat s) \, 
\big|\M_{\LO,Z/\gamma}^{\Pq\bar\Pq\rightarrow \ell\bar \ell}\big|^2, 
\\
2 \Re \Big\{
\M^{\Pq\bar\Pq\rightarrow \ell\bar \ell}_{\text V_\Ps,\text{I}} 
\left( \M_{\text V_\text{p},\text{I}}^{\Pq\bar\Pq\rightarrow \ell\bar \ell} \right)^* 
\Big\} &= 
\delta_{\text{V}_\Ps\otimes\text{V}_\text{p}}^{Z\bar\Pq\Pq,[1 \times 1]}(\hat s)\,  
\big|\M_{\LO,Z/\gamma}^{\Pq\bar\Pq\rightarrow \ell\bar \ell}\big|^2.
\end{align}
We recall that we evaluate these corrections without applying the PA, so that
all amplitudes involve both $\PZ$-boson and photon exchange.
The explicit expressions for the factorized correction factors can be extracted from the 
subleading colour contribution of the $\mathcal{O}(\alphas^2)$ correction to the 
$\Pq\bar\Pq\rightarrow \ell\bar \ell$ amplitude \cite{Gehrmann-DeRidder:2004ttg}
or from the quark formfactor~\cite{Gehrmann:2005pd}
using an abelianization procedure. 
We have recalculated the correction factors explicitly along the same lines as
sketched in \refapp{se:app:formfactors} for the $\QCD \times \text{weak}$ corrections.%
\footnote{The coefficient of the $\epsilon^{-1}$ contribution to $\delta_{\text{V}_\Ps\otimes\text{V}_\text{p}}^{Z\bar\Pq\Pq,[1 \times 1]}$ in \cite{Gehrmann-DeRidder:2004ttg} differs from our result 
by a sign in the term proportional to $\zeta_3$.}
The results are
\begin{align}
\delta_{\text{V}_\Ps\otimes\text{V}_\text{p}}^{Z\bar\Pq\Pq,[2 \times 0]}(\hat s) ={}&
2 Q_q^2 \CF \frac{\alphas\alpha}{\pi^2}\; \mathcal{C}_\epsilon^2\left(\frac{\mu^2}{\hat s}\right)^{2\epsilon}\Bigg[ \frac{1}{4\epsilon^4} + \frac{3}{4\epsilon^3}
+ \frac{1}{\epsilon^2} \left( \frac{41}{16} -\frac{13\pi^2}{24} \right)
 \\
&+ \frac{1}{\epsilon} \left( \frac{221}{32} -\frac{3\pi^2}{2}
-\frac{8}{3}\zeta_3 \right)
+\left(\frac{1151}{64}
          - \frac{475\pi^2}{96}
          - \frac{29}{4}\zeta_3   + \frac{59\pi^4}{288}
   \right)+ {\cal O}(\epsilon) \Bigg]\nn
 ,\\
 \delta_{\text{V}_\Ps\otimes\text{V}_\text{p}}^{Z\bar\Pq\Pq,[1 \times 1]}(\hat s)
 ={}&
2 Q_q^2 \CF \frac{\alphas\alpha}{\pi^2}\;\mathcal{C}_\epsilon^2\left(\frac{\mu^2}{\hat s}\right)^{2\epsilon}
  \Bigg[ \frac{1}{4\epsilon^4} + \frac{3}{4\epsilon^3}
+ \frac{1}{\epsilon^2} \left( \frac{41}{16} -\frac{\pi^2}{24} \right) \label{eq:VsTVphot}\nn\\
&+ \frac{1}{\epsilon} \left( 7 -\frac{\pi^2}{8}
-\frac{7}{6}\zeta_3 \right)
+\left(18
          - \frac{41\pi^2}{96}
          - \frac{7}{2}\zeta_3           - \frac{7\pi^4}{480}
   \right) + {\cal O}(\epsilon)  \Bigg],
\end{align}
where $\mathcal{C}_\epsilon = (4\pi)^\epsilon \, \re^{-\epsilon \gamma_{\mathrm{E}}}$.

\subsubsection{Real--virtual corrections}
\label{se:calc-ii-rv}

The factorizable real--virtual II $\mathcal{O}(\alphas \alpha)$ corrections receive 
contributions from virtual photonic corrections to $\PZ/\gamma^*$~production with real QCD radiation,
from real photon radiation to the virtual QCD corrections to $\PZ/\gamma^*$ production, 
and lastly from virtual weak corrections to $\PZ$~production with real QCD radiation,
as summarized in Fig.\,\ref{fig:rv-diags}. 
The types of
contributing partonic channels of $\PZ+{}$jet and $\PZ+\gamma$ production are given by
\begin{align}
\label{eq:II:rv_channels}
\bar\Pq_a(p_a) + \Pq_b(p_b) &\rightarrow Z(p_Z) + g(k_g),\\
\bar\Pq_a(p_a) + \Pq_b(p_b) &\rightarrow Z(p_Z) + \gamma(k_\gamma),\\
g(p_g)+\Pq_b(p_b) &\rightarrow Z(p_Z)+\Pq_a(k_a),\\
g(p_g)+\bar\Pq_a(p_a) &\rightarrow Z(p_Z)+\bar\Pq_b(k_b).
\end{align}
Note we have not included channels with photons in the initial state, since their 
impact is already suppressed at NLO. 
As described above,
all $\QCD \times \text{photonic}$ II~corrections are consistently evaluated
for the full off-shell process with $\PZ$ and $\gamma$ exchange,
while the amplitudes with a virtual weak and 
a real QCD correction have to be evaluated in \PPA\ 
to preserve gauge invariance.

The $2\to3$ one-loop matrix elements are evaluated in two independent ways.
In the first calculation the results 
for the virtual one-loop EW corrections to $\PZ+{}$jet production have been taken from the earlier 
calculation~\cite{Denner:2011vu}, and the virtual one-loop QCD corrections to $\PZ+\gamma$
have been obtained via an abelianization of the NNLO QCD calculation~\cite{Gehrmann-DeRidder:2023urf}.
In the second calculation \textsc{FeynArts}~\cite{Hahn:2000kx}
is used to generate the amplitudes which are reduced to standard integrals using \textsc{FormCalc}~\cite{hahn1999automated}.
The one-loop integrals are numerically evaluated with the library 
{\sc Collier}~\cite{Denner:2016kdg}.

We now turn to the construction of the PA for the $\QCD \times \text{weak}$ II~corrections,
in particular to the issue of an appropriate OS projection of momenta to
guarantee gauge-invariant subamplitudes.
Having constructed the amplitudes for the virtual weak and real QCD correction 
to Z~production in the $\bar\Pq\Pq$~channel,
we first construct a preliminary version of the PA~amplitude 
from the $\PZ g$-production and \PZ-decay subamplitudes,
\begin{align}
  \label{eq:II:rv_weakxqcd}
    \widetilde\M_{\text V_\Pw\otimes\text R_\Ps,Z,\text{prod}\times\text{prod}}^{\bar\Pq_a \Pq_b \to \Pl \bar\Pl,\PPA} =
  \sum_{\lambda_Z}
  \frac{\M_{\text V_\Pw \otimes \text R_\Ps,\PPA}^{\bar\Pq_a \Pq_b \to gZ}(\lambda_Z)\;
    \M_{0,\PPA}^{Z\to\Pl\bar\Pl}(\lambda_Z)}{p_Z^2 - \mu_Z^2},
\end{align}
where the tilde indicates that we still have to fix the OS projection of the external
momenta.
Rescaling all external momenta according to
\begin{align}
 p_i \to{}& \hat p_i = p_i \, \frac{\MZ}{\sqrt{2 k_\ell k_{\bar\ell}}},\qquad i=a,b,g,
\nn\\
 k_j \to{}& \hat k_j = k_j \, \frac{\MZ}{\sqrt{2 k_\ell k_{\bar\ell}}},\qquad j=a,b,\ell,\bar\ell,g,
 \label{eq:os-mapping-1}
\end{align}
preserves on-shellness of all (light-like) momenta and momentum conservation, and forces
the new $\PZ$~momentum $\hat p_Z = \hat k_\ell+\hat k_{\bar\ell}$ on-shell,
\begin{align}
 \hat p_Z^2 = 2 \, \hat k_\ell \hat k_{\bar\ell} = \MZ^2.
\end{align}
Simply applying Eq.~\refeq{eq:os-mapping-1} to the residue of the resonance in
Eq.~\refeq{eq:II:rv_weakxqcd} is, however,
not sufficient to define a consistent PA, since we also have to guarantee a proper
subtraction of all soft and collinear divergences in the IR limits. Since the
subtraction function is constructed from the underlying $2\to2$ scattering amplitudes
in PA, the OS projections used in the $2\to2$ and $2\to3$ contributions are not
independent.

In the calculation of the double-virtual corrections described in the previous section
we have defined the OS projection in such a way that only the dimensionless 
$\PZ\bar\Pq\Pq$ formfactor is forced to be on-shell, but not the Dirac chains 
${\cal A}^{\sigma\tau}$
of the amplitudes (and of course not the $\PZ$~propagator containing the resonance).
This variant ensures the same energy-scaling behaviour of the PA and off-shell 
amplitudes in the far off-shell regions up to logarithmic deviations 
contained in the corrections. Breaking the scaling behaviour would be prone to artefacts
when evaluating the PA 
on the full phase space.
The OS projection of the $2\to2$ amplitudes can be summarized as follows,
\begin{align}
 \M_{Z,\PPA}^{2\to2}(p_a,p_b,k_\ell,k_{\bar\ell}) =  
\frac{\big[\M_{Z}^{2\to2}\cdot (p_Z^2-\mu_Z^2) 
\big]\big\vert_{p_i \to \hat p_i,k_j\to \hat k_j}}{p_Z^2-\mu_Z^2}  
\cdot \frac{2\, k_\ell k_{\bar\ell}}{M_Z^2},
 \label{eq:scaling-2to2-PA}
\end{align}
where the last factor on the r.h.s.\ is used to restore the original
scaling behaviour of the $\mathcal{A}^{\sigma\tau}$ after the application of \refeq{eq:os-mapping-1}.

By the same reasoning, we have to rescale $\widetilde\M_{\text V_\Pw\otimes\text R_\Ps,Z,\text{prod}\times\text{prod}}^{\bar\Pq_a \Pq_b \to \Pl \bar\Pl,\PPA}$ of Eq.~\refeq{eq:II:rv_weakxqcd}
by a factor $2\, k_\ell k_{\bar\ell}/M_Z^2$ to restore the scaling behaviour
of the spinor chains. However, we have yet to apply another factor
$\MZ/\sqrt{2\, k_\ell k_{\bar\ell}}$ to compensate for the fact that the $2\to3$
amplitudes contain another factor of $(\mbox{energy})^{-1}$ that was rescaled
by applying Eq.~\refeq{eq:os-mapping-1}. If we did not include this compensation
factor, some mismatch would arise with the OS-projected subtraction function
which employs squared $2\to2$ amplitudes in PA times a splitting factor
of dimension $(\mbox{energy})^{-2}$ that is based on the off-shell kinematics.
In summary, the OS projection of the $2\to3$ amplitudes reads
\begin{align}
\M_{\text V_\Pw\otimes\text R_\Ps,Z,\text{prod}\times\text{prod}}^{\bar\Pq_a \Pq_b \to \Pl \bar\Pl,\PPA} = 
\frac{\big[\widetilde\M_{\text V_\Pw\otimes\text R_\Ps,Z,\text{prod}\times\text{prod}}^{\bar\Pq_a \Pq_b \to \Pl \bar\Pl,\PPA} \cdot (p_Z^2-\mu_Z^2) \big]\big\vert_{p_i \to \hat p_i,k_j\to\hat k_j}}{p_Z^2-\mu_Z^2}  \cdot \sqrt{\frac{2\, k_\ell k_{\bar\ell}}{M_Z^2}}.
\end{align}
This OS projection of the $2\to3$ amplitudes is also in line with the integrated 
subtraction function which receives the same scaling as the $2\to2$ contributions
after integration over the off-shell phase space.

Note that the choice we made for the OS projection is not unique 
and different choices would be possible. 
For instance, we could have opted to keep also the Dirac structures on-shell. 
Different versions for the OS projection lead to results that formally differ at the order 
of the intrinsic uncertainty of the PA. 
However, it is important to construct the on-shell projection in a self-consistent way 
in order to guarantee a proper subtraction of all soft and collinear singularities 
between  $2\to 3$ contributions and subtraction functions in the
IR limits.
The OS projection for $2\to 3$ processes, thus,  implicitly fixes 
OS projection of all $2\to 2$ processes by this consistency requirement.

\subsubsection{Double-real corrections}
\label{se:calc-ii-rr}

The double-real corrections are diagrammatically illustrated
in \reffi{fig:rr-diags} and are induced by diagrams with both an external gluon and photon
or by diagrams with two additional external (anti)quarks and an internal photon,
i.e.\ they are all part of the $\QCD \times \text{photonic}$ 
$\mathcal{O}(\alphas \allowbreak \alpha_\text{p})$ II~corrections,
while QCD$\times$weak corrections do not contribute here.%
\footnote{%
We note that the last diagram in \reffi{fig:rr-diags} has a weak counterpart 
(with a Z~boson instead of the photon in the blob)
that is expected to be strongly suppressed and thus not considered further.}
Accordingly, we do not apply the PA in the calculation of double-real corrections. 
The types of
channels that have to be considered are given by
\begin{align}
\label{eq:II:rr_channels}
\bar\Pq_a(p_a) + \Pq_b(p_b) &\rightarrow Z(p_Z) + g(k_g)+\gamma(p_\gamma),\\
g(p_g)+\Pq_b(p_b) &\rightarrow Z(p_Z)+\Pq_a(k_a)+\gamma(p_\gamma),\\
g(p_g)+\bar\Pq_a(p_a) &\rightarrow Z(p_Z)+\bar\Pq_b(k_b)+\gamma(p_\gamma),\\
\Pq_b(p_a)+\Pq_b(p_b) &\rightarrow Z(p_Z) +\Pq_b(k_a)+\Pq_b(k_b),\\
\bar\Pq_a(p_a)+\bar\Pq_a(p_b) &\rightarrow Z(p_Z) +\bar\Pq_a(k_a)+\bar\Pq_a(k_b),\\
\bar\Pq_a(p_a)+\Pq_a(p_b) &\rightarrow Z(p_Z) +\bar\Pq_a(k_a)+\Pq_a(k_b).
\end{align}
Again, we have not included channels with photons in the initial state, since their 
impact is already suppressed at NLO. 
The helicity amplitudes for the considered partonic channels were calculated using 
the spinor-helicity formalism, using the formulation of \citere{Dittmaier:1998nn},
and independently through the abelianization of the corresponding NNLO QCD amplitudes.
Note that the double-real correction induced by the last diagram of \reffi{fig:rr-diags} 
is only non-zero for the case where the quark chains close to a single loop, i.e.\
for the scattering of two identical quarks. Otherwise the interference amplitude
vanishes owing to colour conservation.

\section{Numerical results}
\label{se:num-res}

\subsection{Input parameters and event selection}
\label{se:setup}

The setup for the calculation is largely taken over from
Refs.~\cite{Dittmaier:2014qza,Dittmaier:2015rxo}.
The choice of input parameters closely follows
Ref.~\cite{Beringer:1900zz},
\begin{equation}
\label{eq:params}
\begin{aligned}
  M_{\PW,\OS} \;=&\; 80.385 \GeV ,
  &\Gamma_{\mathrm W,\OS} \;=&\; 2.085 \GeV , \\
  M_{\PZ,\OS} \;=&\; 91.1876 \GeV ,
  &\Gamma_{\PZ,\OS} \;=&\; 2.4952 \GeV , \\
  M_\PH \;=&\; 125.9 \GeV , 
  &G_\mu \;=&\; 1.1663787\times 10^{-5}  \GeV^{-2} , \\
  m_\Pt \;=&\; 173.07 \GeV , 
  &m_\Pb \;=&\; 4.78\GeV , \\
  m_\mu \;=&\; 105.658369\MeV.
%\\
%  \alphas(\MZ) \;=&\; 0.119 .
\end{aligned}
\end{equation}
We convert the on-shell (OS) masses
and decay widths of the vector bosons to the corresponding pole masses
according to~\cite{Denner:2019vbn}
\begin{equation}
M_V = \frac{M_{V,\OS}}{\sqrt{1+\Ga_{V,\OS}^2/M_{V,\OS}^2}},
\qquad
\Gamma_V = \frac{\Ga_{V,\OS}}{\sqrt{1+\Ga_{V,\OS}^2/M_{V,\OS}^2}}.
\end{equation}
The electromagnetic coupling constant is set according to the 
$G_\mu$ scheme.
The masses of the light quark flavours (u,d,c,s) 
and of the leptons are neglected throughout. 
The mass $m_\mu$ of the muon is only needed in the evaluation of the 
logarithmically mass-singular FSR corrections for bare muons.
The CKM matrix is chosen diagonal in the third generation, 
and the mixing between the first two generations of massless quarks cancels in
cross sections by virtue of the unitarity of the CKM matrix.
%is parametrized by the following values for the entries of the quark-mixing matrix,
%\begin{equation}
%  \label{eq:ckm}
%  \lvert V_{\Pu\Pd} \rvert \,=\, 
%  \lvert V_{\Pc\Ps} \rvert \,=\, 0.974, \qquad
%  \lvert V_{\Pc\Pd} \rvert \,=\, 
%  \lvert V_{\Pu\Ps} \rvert \,=\, 0.227. 
%\end{equation}
While b-quarks appearing in closed fermion
loops have the mass $m_\Pb$ given in Eq.~(\ref{eq:params}), external b-quarks are taken as massless.
 
We consider $\mu^-\mu^+$ production in pp collisions at a centre-of mass energy of
$13\TeV$.
For the PDFs we consistently use the NNPDF3.1 set~\cite{Bertone:2017bme}, 
i.e.\ the NLO and NNLO QCD and
QCD${}\times{}$EW corrections are evaluated using the
\verb|NNPDF31_nlo_as_0118_luxqed|
set, which also includes $\mathcal{O}(\alpha)$ corrections.
The value of the strong coupling $\alphas(\MZ)=0.118$ is dictated by the choice of this PDF set.
The renormalization and factorization scales are set equal, with a fixed value given by 
the $\PZ$-boson mass,
\begin{equation}
  \label{eq:scale}
  \mu_{\mathrm{R}} \;=\; \mu_{\mathrm{F}} \;=\; \MZ .
\end{equation}

For the experimental identification of the 
DY process we impose the following cuts on the transverse momenta and rapidities of the charged leptons,
\begin{align}
  \label{eq:cut-lep}
  k_{\rT,\Pl^\pm} > 25\GeV , \qquad
  \lvert y_{\Pl^\pm} \rvert < 2.5 . 
\end{align}
We further apply a cut on the invariant mass $M_{\Pl\Pl}$ of the lepton pair,
\begin{equation}
  \label{eq:cut-mll}
  M_{\Pl\Pl} > 50\GeV ,
\end{equation}
in order to avoid the photon pole at $M_{\Pl\Pl}\to0$.

In the following, we distinguish two
alternative treatments of photon radiation off leptons:
``bare muons'' and ``dressed leptons''.
In the bare-muon case, no recombination of leptons and 
nearly collinear photons is performed, reflecting the experimental situation
which allows for the detection of isolated muons.
In the dressed-lepton
case, collinear photon--lepton configurations are treated inclusively
using a photon-recombination procedure.  As a result, the numerical
predictions do not contain large logarithms of the lepton mass, which
can be set to zero. The dressed-lepton results are appropriate
mostly for electrons in the final state. 
In detail, for dressed leptons a photon recombination procedure analogous to the 
one used in Refs.~\cite{Dittmaier:2001ay,Dittmaier:2009cr} is applied:
\begin{enumerate}
\item Photons close to the beam with a rapidity $\lvert \eta_{\gamma} \rvert > 3$ are treated as beam remnants and are not further considered in the event selection.
\item For the photons that pass the first step, the angular distance to the charged leptons 
$R_{\Pl^\pm\gamma}=\sqrt{(\eta_{\Pl^\pm}-\eta_{\gamma})^2 + \Delta\phi_{\Pl^\pm\gamma}^2}$ 
is computed, where $\Delta\phi_{\Pl^\pm\gamma}$ 
denotes the azimuthal angle difference
of the lepton and photon in the transverse plane.
  If the distance $R_{\Pl^\pm\gamma}$
between the photon and the closest lepton is smaller than
$R_{\Pl^\pm\gamma}^{\mathrm{rec}}=0.1$, the photon is recombined with the lepton by adding the respective four-momenta, 
$\Pl^\pm(k_i)+\gamma(k)\to\Pl^\pm(k_i+k)$.
\item Finally, the event selection cuts from Eqs.~\eqref{eq:cut-lep}--\eqref{eq:cut-mll} are applied to the resulting event kinematics.
\end{enumerate}

\subsection{Corrections to differential distributions}

In this section we discuss our results on the corrections to muon pair production
($\ell=\mu$) and to the production of a dressed lepton pair 
for the distributions in the 
invariant mass $M_{\ell\ell}$ of the $\bar\ell\ell$ pair, 
the transverse momentum $k_{\rT,\ell}$ of one of the leptons, 
and the rapidity $y_{\ell\ell}$ of the $\bar\ell\ell$ pair.
By default, relative corrections $\delta$ are normalized to full off-shell LO distributions
in the following.
Whenever of relevance, we show both results for the event selection without
or with  photon recombination, denoted
``bare muons'' and ``dressed leptons'', respectively, in the following.

\begin{figure}
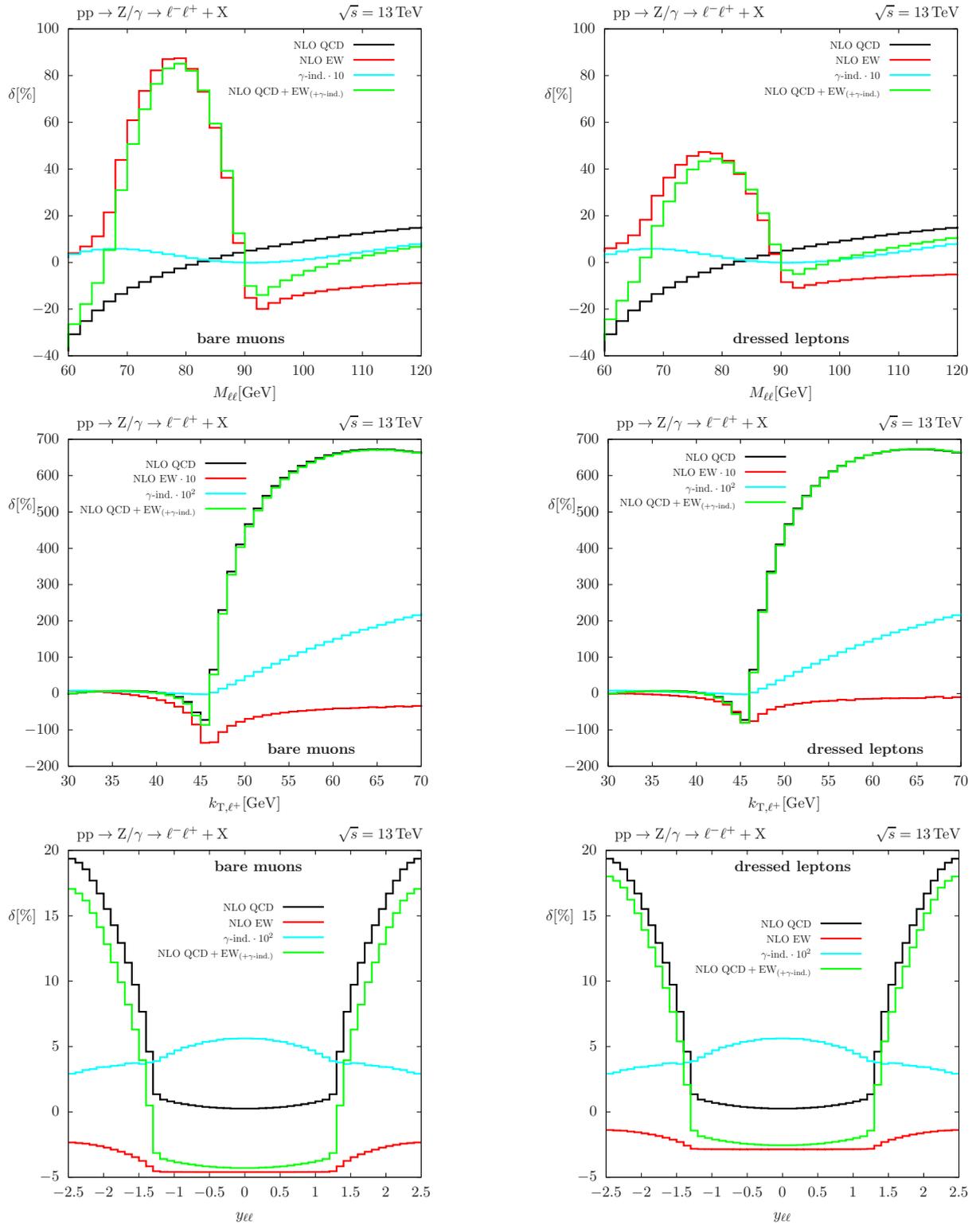

\includegraphics[scale=.65]{{{plots/NLO/Z-low.mll.NLO.bare}}}
\hfill
\includegraphics[scale=.65]{{{plots/NLO/Z-low.mll.NLO.dressed}}}
\\
\includegraphics[scale=.65]{{{plots/NLO/Z-low.ktl1.NLO.bare}}}
\hfill
\includegraphics[scale=.65]{{{plots/NLO/Z-low.ktl1.NLO.dressed}}}
\\
\includegraphics[scale=.65]{{{plots/NLO/Z-low.yll.NLO.bare}}}
\hfill
\includegraphics[scale=.65]{{{plots/NLO/Z-low.yll.NLO.dressed}}}
\caption{Relative NLO QCD (black), NLO EW (red),
$\gamma$-induced (sum of LO $\gamma\gamma$ and NLO $q\gamma/\bar q\gamma$ contributions in cyan),
and full NLO (green) corrections (normalized to LO)
to various distributions for bare muons (l.h.s.) and for dressed leptons (r.h.s.).}
\label{fig:NLOdist}
\end{figure}

We start out by showing the NLO QCD and NLO EW corrections
in \reffi{fig:NLOdist}, recalculated in the
setup described in the previous section. 
\begin{figure}
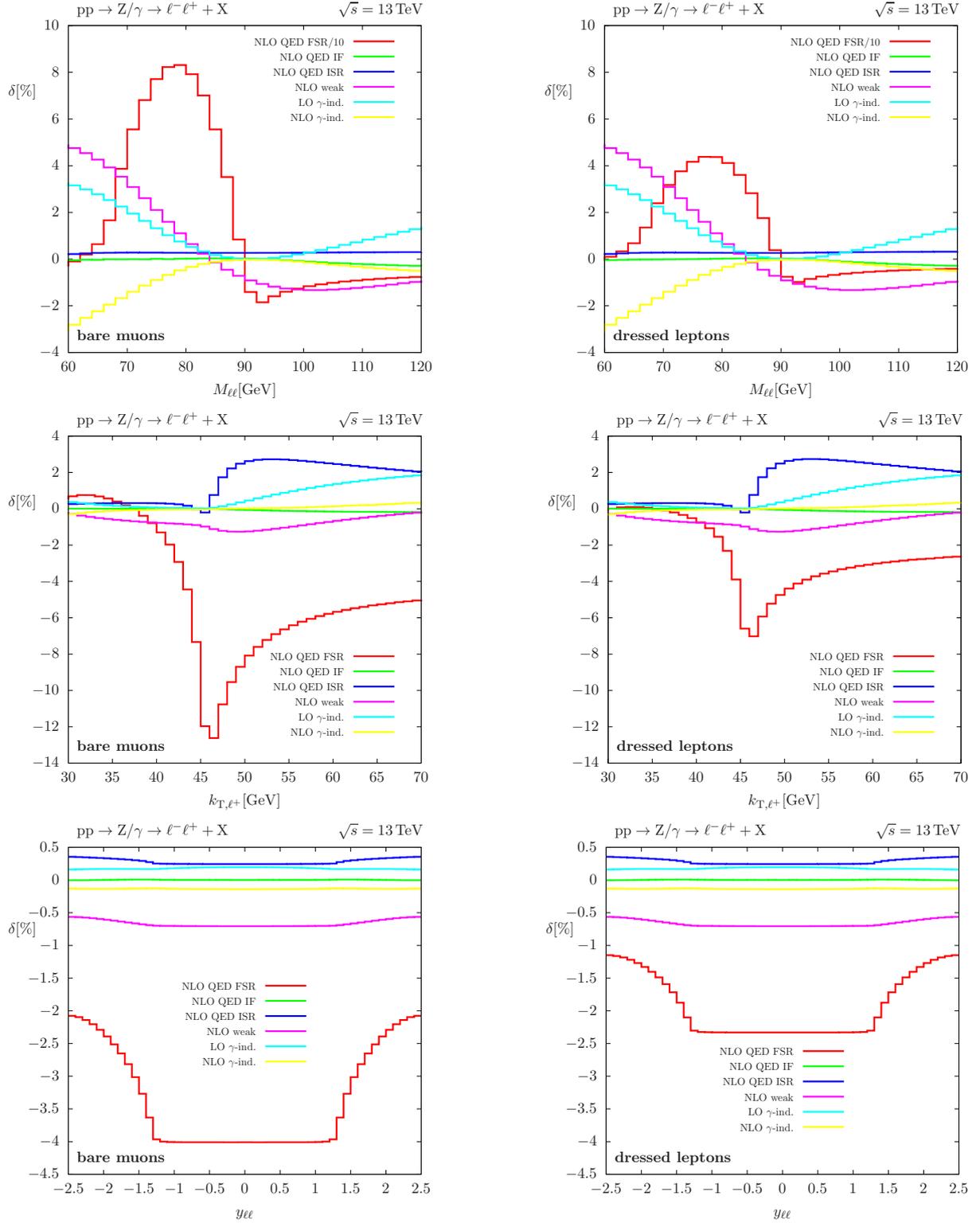

\includegraphics[scale=.65]{{{plots/NLO_EW/Z-low.mll.NLO_EW.bare}}}
\hfill
\includegraphics[scale=.65]{{{plots/NLO_EW/Z-low.mll.NLO_EW.dressed}}}
\\
\includegraphics[scale=.65]{{{plots/NLO_EW/Z-low.ktl1.NLO_EW.bare}}}
\hfill
\includegraphics[scale=.65]{{{plots/NLO_EW/Z-low.ktl1.NLO_EW.dressed}}}
\\
\includegraphics[scale=.65]{{{plots/NLO_EW/Z-low.yll.NLO_EW.bare}}}
\hfill
\includegraphics[scale=.65]{{{plots/NLO_EW/Z-low.yll.NLO_EW.dressed}}}
\caption{Decomposition of the NLO EW corrections (normalized to LO)
into photonic FSR (red), 
photonic IF interference effects (green), 
photonic ISR (blue), and 
weak corrections (magenta),
as well as relative corrections from LO $\gamma\gamma$ (cyan)
and NLO $q\gamma/\bar q\gamma$ (yellow)
initial states.}
\label{fig:NLOdistEWdecomp}
\end{figure}
In the following,
all LO, NLO, and NNLO cross sections are evaluated with the same PDF set, in order
to make the impact of all types of corrections most transparent, without
being affected by differences in the PDFs.
The LO and NLO predictions are calculated using the 
complex-mass scheme~\cite{Denner:2005fg,Denner:2019vbn} for treating the
$\PZ$-boson resonance, as described in \citere{Dittmaier:2009cr} in detail,
and the PA is not used in any of the NLO corrections.
Owing to the suppression of the contribution from $\gamma\gamma$ initial
states, we only include the contributions of the $q\bar q$ channels in the LO cross section
and consider the LO $\gamma\gamma$ contribution as part of the corrections.
We denote as ``$\gamma$-ind.'' the sum of the LO ($\gamma\gamma$) and
NLO ($q\gamma/\bar q\gamma$) photon-induced contributions.
As will be illustrated in the following, the photon-induced contributions are strongly suppressed with no particular enhancement mechanism such that corrections beyond NLO, i.e.\ at $\mathcal{O}(\alphas\alpha)$, can be safely neglected for all phenomenological purposes.%
\footnote{The omission of higher-order corrections to both the di-photon and other photon-induced contributions is further supported by Ref.~\cite{Bonciani:2021zzf}, see e.g.\ Table~I.}

The decomposition of the NLO EW corrections into contributions from
photonic final-state radiation (FSR), 
photonic initial--final interference effects (IF), 
photonic initial-state radiation (ISR), and 
purely weak corrections (weak) are shown in \reffi{fig:NLOdistEWdecomp}.
The LO $\gamma\gamma$ contribution as well as the NLO
correction induced by $q\gamma/\bar q\gamma$ initial states are illustrated
in \reffi{fig:NLOdistEWdecomp} separately as well.
Of course, similar results have been shown in
several places in the literature (see, e.g., \citere{Dittmaier:2009cr}), 
but recalling these results facilitates the discussion of the
$\mathcal{O}(\alphas\alpha)$ corrections below.

The EW corrections to the $M_{\ell\ell}$ distributions show the well-known radiative
tail below the resonance at $M_{\ell\ell}=\MZ$, which is dramatically enhanced due to
the collinear singularity $\propto\alpha\ln(m_\ell/\MZ)$ in the FSR contribution
if no photon recombination is applied. 
The fact that the radiative tail does not extend below $M_{\ell\ell}$ values of about
$68\GeV$ is an effect of the acceptance cuts on the lepton transverse momenta
$k_{\rT,\Pl^\pm}$ of $25\GeV$. The major part of all Z~bosons is produced on shell,
so that before FSR most leptons carry $k_{\rT,\Pl^\pm}$ of at most $\MZ/2$.
At NLO, collinear FSR reduces one of the lepton momenta by a factor $z$ ($0<z<1$),
so that $M_{\ell\ell}^2$ is given by $z \MZ^2$ after FSR.
The maximal value of the reduced $k_{\rT,\Pl^\pm}$ is $z\MZ/2$, corrsponding
to a Z-boson decay transverse to the beams. Thus, for $k_{\rT,\Pl^\pm}=z\MZ/2<25\GeV$, 
which corresponds to $M_{\ell\ell}=\sqrt{z}\MZ<68\GeV$,
events with an on-shell Z~boson and collinear FSR off a lepton cannot pass the acceptance
cut any more, which leads to a strong suppression of the FSR correction
for such invariant masses $M_{\ell\ell}$.
The photon recombination described in the previous section mitigates
the FSR corrections by roughly a factor~2,
i.e.\ the mass-singular logarithm is effectively replaced by
$\sim\ln R_{\Pl^\pm\gamma}^{\mathrm{rec}}=\ln(0.1)$.
The remaining photonic corrections (IF, ISR) are at the sub-percent level,
the weak corrections at the few-percent level.
The corrections induced by $\gamma\gamma$ and $q\gamma/\bar q\gamma$ initial states
are largely suppressed on resonance (the $\gamma\gamma$ channel does not
develop a Z~resonance at all), but typically matter at the percent level in a
window of a width of $10{-}20\GeV$ around the Z-boson resonance.

The distribution in the transverse momentum of a lepton at LO is dominated by
resonant $\PZ$~production for $k_{\rT,\ell}\lesssim\MZ/2$, where all NLO corrections
are moderate; the largest EW effects are again due to FSR.
For larger $k_{\rT,\ell}\gtrsim\MZ/2$ the QCD corrections develop
``giant $K$-factors''~\cite{Rubin:2010xp},
since the jet recoil in the real QCD corrections allows for 
the population of the region with $k_{\rT,\ell}>\MZ/2$ by events with resonant
$\PZ$~bosons.
The $\gamma\gamma$- and $q\gamma/\bar q\gamma$-induced corrections only amount to
some $\sim 0.1\%$ for transverse momenta below the Jacobian peak.
For large transverse momenta, the same is true for the relative corrections
normalized to the full (QCD-corrected) differential cross section.

Finally, the NLO corrections to the rapidity distribution of the $\PZ$~boson resemble
the moderate corrections to the integrated cross section in the central part
of the distribution, i.e.\ for $|y_{\ell\ell}|\lesssim1.3$.
The corrections from $\gamma\gamma$ and $q\gamma/\bar q\gamma$ initial states
only contribute at the level of $0.1{-}0.2\%$ over the whole rapidity range.

\begin{figure}
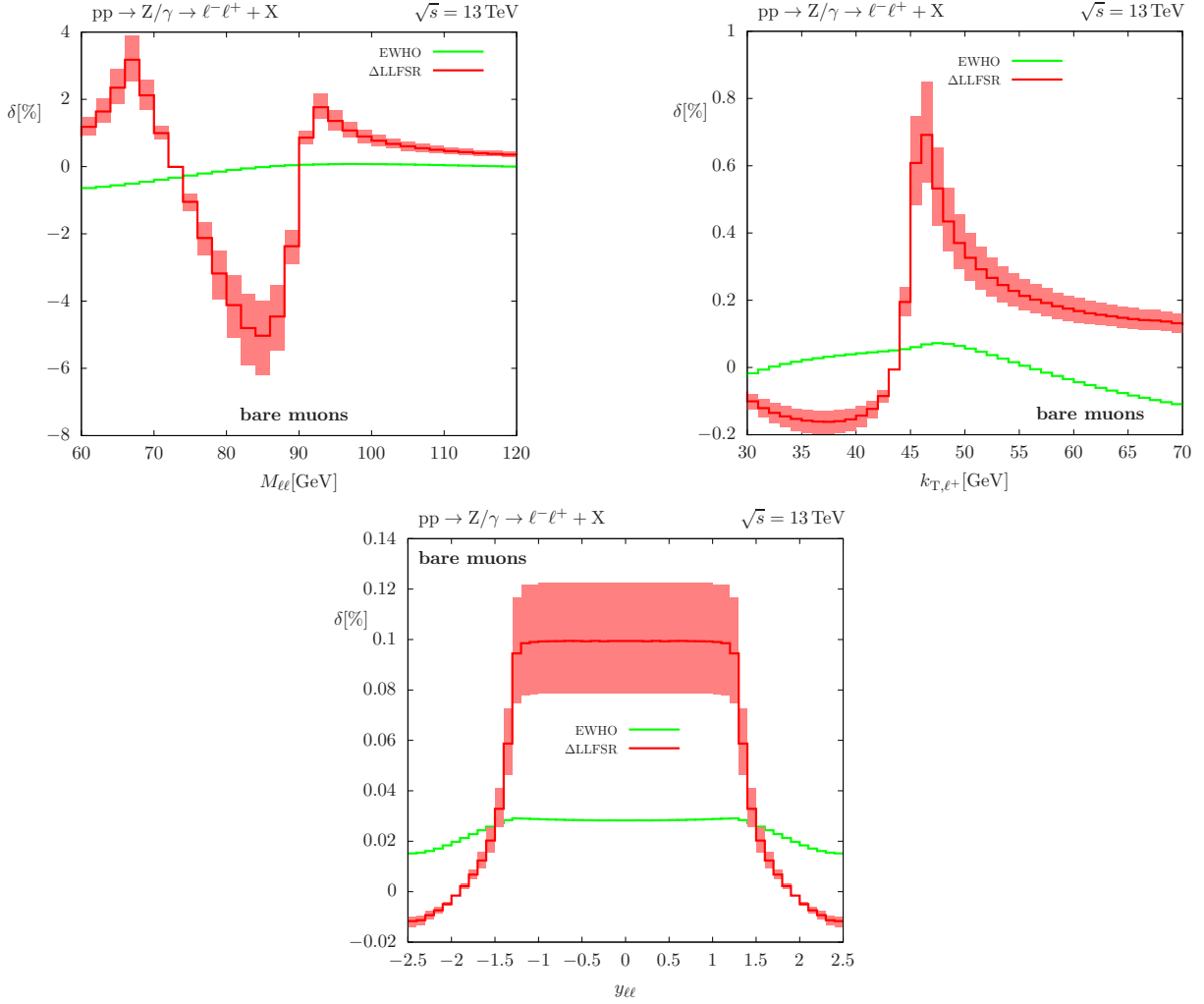

\includegraphics[scale=.65]{{{plots/LLFSR_EWHO/Z-low.mll.EWHO_LLFSR.bare}}}
\hfill
\includegraphics[scale=.65]{{{plots/LLFSR_EWHO/Z-low.ktl1.EWHO_LLFSR.bare}}}
\hfill
\centering
\includegraphics[scale=.65]{{{plots/LLFSR_EWHO/Z-low.yll.EWHO_LLFSR.bare}}}
\hfill
\caption{Relative higher-order photonic FSR effects ($\Delta$LLFSR, red),
i.e.\ the NLO contribution is subtracted, 
and universal EW higher-order corrections (EWHO, green), both normalized to LO.
The red band illustrates the scale uncertainty of the LLFSR correction by varying the
central FSR scale $\mu_\text{FSR}=\MZ$ up and down by a factor~2.}
\label{fig:hoEWdist}
\end{figure}

Figure~\ref{fig:hoEWdist} shows two types of higher-order EW corrections
beyond NLO:
First, the FSR effects induced by collinear multi-photon emission off
the leptons in the structure function 
approach~\cite{Kuraev:1985hb,Nicrosini:1986sm,Nicrosini:1987sw,Berends:1987ab,Arbuzov:1999cq,Blumlein:2007kx}%
\footnote{
  Note that Ref.~\cite{Berends:1987ab} contains errors that have been addressed in Refs.~\cite{Blumlein:2011mi,Blumlein:2020jrf}.
}
based on leading logarithms up to $\mathcal{O}(\alpha^3)$ 
with the NLO contribution subtracted.
Second, the leading NNLO EW effects from the universal
corrections induced by the running of the electromagnetic coupling ($\Delta\alpha$) 
and by corrections to the $\rho$-parameter ($\Delta\rho$). 
The precise definition of the two types of corrections can be found in
Sects.~3.4.3 and 3.4.1 of \citere{Dittmaier:2009cr}, respectively.
The corrections shown in \reffi{fig:hoEWdist} are obtained for bare muons.
For dressed leptons the FSR effects based on leading-log structure functions vanish,
and the leading universal EW corrections would be identical to the ones for
bare muons, because these corrections do not involve photon radiation.
The most notable higher-order EW effect shown in \reffi{fig:hoEWdist}
arises from multi-photon correction
to the $M_{\ell\ell}$ distribution, with an impact on the radiative tail at
the level of $5\%$. Note also that the uncertainty arising from the scale
$\mu_\text{FSR}$ of the multi-photon effects that is not unambiguously
fixed in leading logarithmic approximation 
is not completely 
negligible in the $M_{\ell\ell}$ distribution.

\begin{figure}
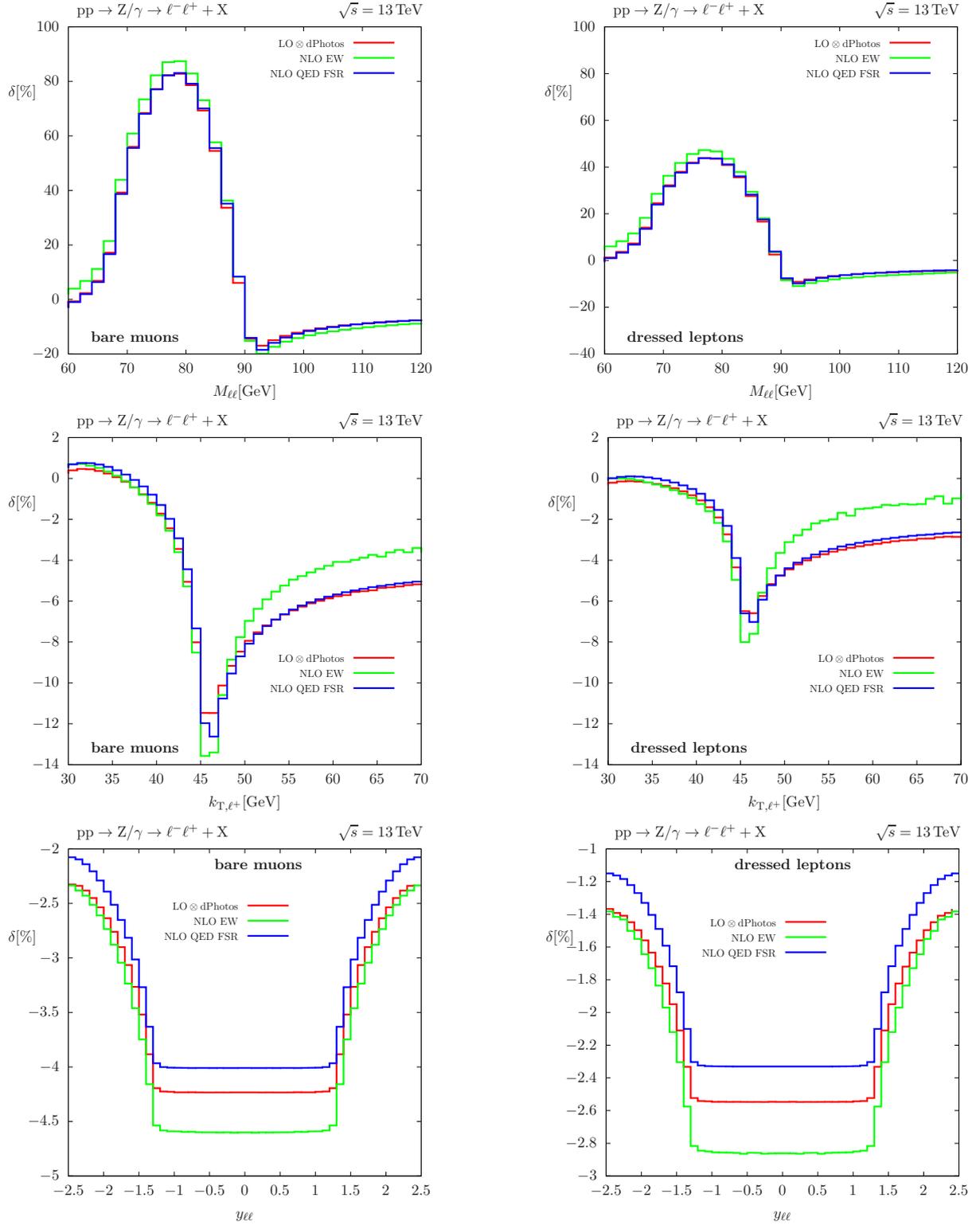

\includegraphics[scale=.65]{{{plots/NLO_photos/Z-low.mll.NLO.photos.bare}}}
\hfill
\includegraphics[scale=.65]{{{plots/NLO_photos/Z-low.mll.NLO.photos.dressed}}}
\includegraphics[scale=.65]{{{plots/NLO_photos/Z-low.ktl1.NLO.photos.bare}}}
\hfill
\includegraphics[scale=.65]{{{plots/NLO_photos/Z-low.ktl1.NLO.photos.dressed}}}
\includegraphics[scale=.65]{{{plots/NLO_photos/Z-low.yll.NLO.photos.bare}}}
\hfill
\includegraphics[scale=.65]{{{plots/NLO_photos/Z-low.yll.NLO.photos.dressed}}}
\caption{NLO $\text{EW}$ (green) and $\text{LO}_\text{QCD}\otimes\text{dPhotos}$ (red)
corrections, normalized to LO.}
\label{fig:NLO-photos}
\end{figure}
Given that the dominant EW corrections arise from FSR corrections, we compare in Fig.~\ref{fig:NLO-photos} the NLO prediction to the one obtained using the \textsc{Photos}~\cite{Barberio:1993qi}
QED shower on top of the LO prediction.
This allows to assess the performance of a tool commonly employed in the experimental measurements and the potential impact of multi-photon emissions that go beyond the structure-function approach discussed above.
As anticipated, FSR effects are well captured by the \textsc{Photos} tool, agreeing very well with the NLO QED FSR part of our calculation.
NLO EW effects that go beyond FSR, i.e.\ ISR and initial--final QED effects and genuine weak corrections, on the other hand, are not included and thus not captured by the naive LO$\otimes$\textsc{Photos} prediction.
Small differences in the normalisation, as can be seen in the $y_{\ell\ell}$ distribution can likely be attributed to multi-photon effects and the choice of $\alpha$ scheme within the QED shower.

\begin{figure}
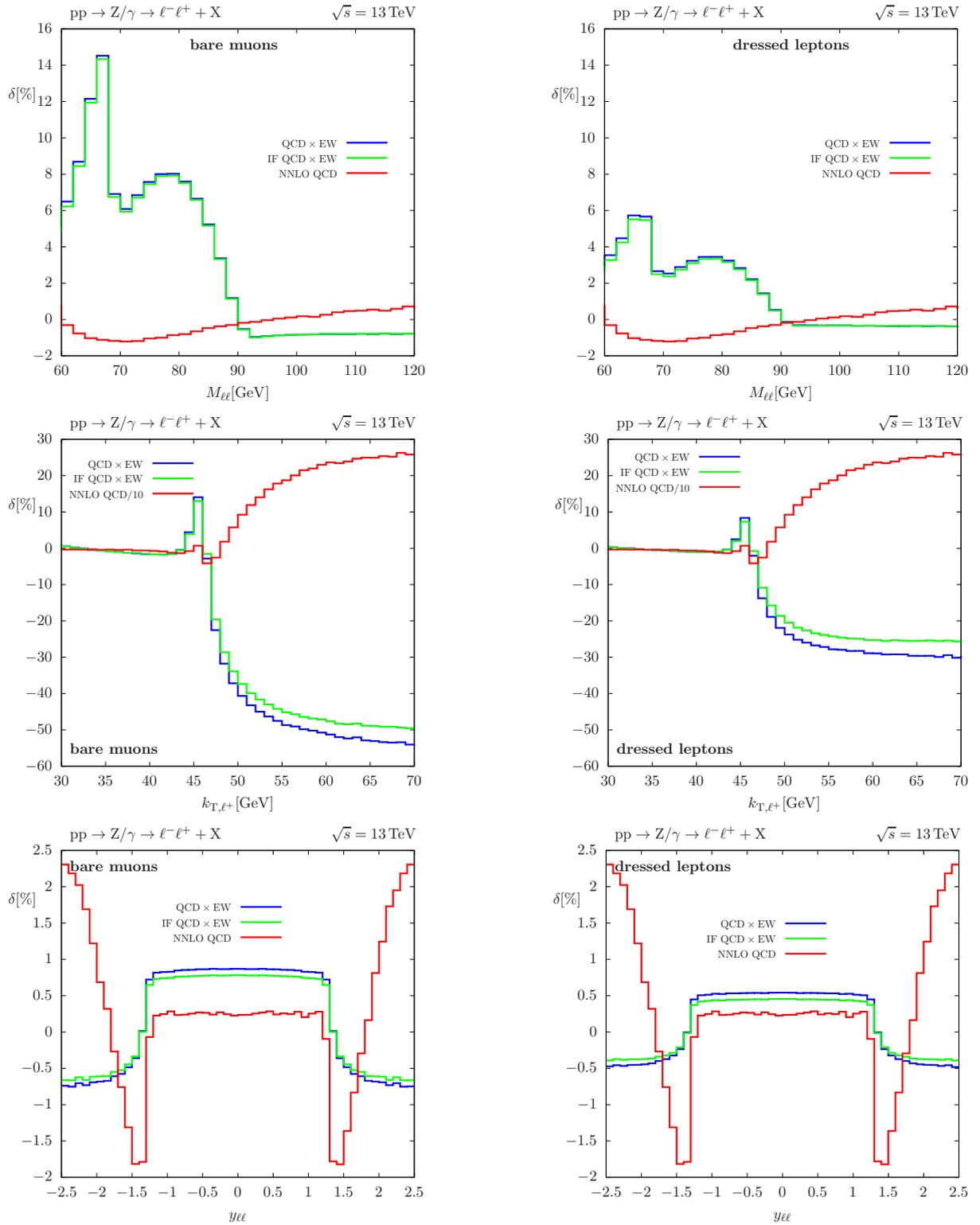

\includegraphics[scale=.65]{{{plots/NNLO/Z-low.mll.NNLO.bare}}}
\hfill
\includegraphics[scale=.65]{{{plots/NNLO/Z-low.mll.NNLO.dressed}}}
\includegraphics[scale=.65]{{{plots/NNLO/Z-low.ktl1.NNLO.bare}}}
\hfill
\includegraphics[scale=.65]{{{plots/NNLO/Z-low.ktl1.NNLO.dressed}}}
\includegraphics[scale=.65]{{{plots/NNLO/Z-low.yll.NNLO.bare}}}
\hfill
\includegraphics[scale=.65]{{{plots/NNLO/Z-low.yll.NNLO.dressed}}}
\caption{Relative NNLO QCD corrections (red) as well as a comparison of the full
$\text{QCD}\times\text{EW}$ corrections in PA (blue) with its IF
contribution (green); all corrections are normalized to LO.}
\label{fig:NNLOdist}
\end{figure}

The NNLO $\text{QCD}\times\text{EW}$ corrections in PA are shown in 
\reffi{fig:NNLOdist} together with their by far dominating contribution 
of type~IF, which was already calculated in \citere{Dittmaier:2015rxo}.
In the regions that are dominated by the $\PZ$~resonance, the 
IF $\text{QCD}\times\text{EW}$ corrections are typically one to a few percent
and thus phenomenologically important and even larger than the NNLO QCD
corrections, which we have been evaluated using the results of 
\citere{Gehrmann-DeRidder:2023urf}.
Note that the large radiative tail for $M_{\ell\ell}\lesssim\MZ$, where the
$\text{QCD}\times\text{EW}$ corrections grow to $\sim5{-}10\%$,
are still calculable in PA, because these effects are dominated by
photonic FSR off the leptons that result from nearly resonant $\PZ$~bosons.
The shape of the $\text{QCD}\times\text{EW}$ correction to the
$M_{\ell\ell}$ distribution is widely inherited from the product of the
photonic FSR effect at $M_{\ell\ell}$ and the QCD ISR correction
at $M_{\ell\ell}\approx\MZ$.
The only exception is the narrow peak slightly below $M_{\ell\ell}\sim68\GeV$,
which is a fixed-order artefact from soft-gluon emission.
To understand the origin of this peak, recall the explanation of the truncation
of the FSR radiative tail observed in \reffi{fig:NLOdist}
below the Z~resonance at $M_{\ell\ell}\sim68\GeV$. For smaller $M_{\ell\ell}$
one of the decay leptons of a resonant Z~boson cannot pass the cuts on
$k_{\rT,\Pl^\pm}$ at NLO. 
At NNLO $\text{QCD}\times\text{EW}$, jet emission before the formation of the Z~resonance
leads to a recoil of the Z~boson that is transferred to the decay leptons.
Thus, near the edge at $M_{\ell\ell}\sim68\GeV$, relatively soft gluon emission
can be enough to allow an event to pass the $k_{\rT,\Pl^\pm}$ cuts while corresponding
events with virtual gluon exchange and LO kinematics in the Z~production process 
still do not pass the cuts. This mismatch leads to the sharp peak for
$M_{\ell\ell}$ values slightly below the edge of the radiative QED FSR tail.
Soft-gluon resummation or possibly an adjustment of fiducial cuts
or of the event selection
would largely mitigate this artefact, but this is beyond
the scope of this paper.

As mentioned in the introduction, the setup of our calculation coincides with the one used
in \citere{Bonciani:2021zzf}, where the $\mathcal{O}(\alphas\alpha)$ corrections
have been evaluated for bare muons without applying the PA.
Overlaying our relative corrections to the $M_{\ell\ell}$ distribution
shown in the top--left plot of \reffi{fig:NNLOdist} with the corresponding plot
in Fig.~2 of \citere{Bonciani:2021zzf} reveals agreement within 
statistical fluctuation.%
\footnote{In the invariant-mass window $70\GeV<M_{\ell\ell}<110\GeV$
the good agreement holds, even though our PA takes into account only
Z-boson exchange LO diagrams, i.e.\ we do not reweight the full LO 
cross section with Z-boson and photon exchange by some PA correction
factor, as suggested in Eqs.~(12,13) of \citere{Buonocore:2021rxx}.}
Note that this agreement also confirms our expectation that 
$\mathcal{O}(\alphas\alpha)$ corrections induced by photons in the
initial state, which are neglected in our calculation but included in the results
of \citere{Bonciani:2021zzf}, are negligible.

In the $k_{\rT,\ell}$ distribution the only significant effect of $\sim10\%$
appears at $k_{\rT,\ell}\lesssim\MZ/2$ where the region of resonant $\PZ$~bosons
sets in. Note that for $k_{\rT,\ell}>\MZ/2$ the plots overestimate the
impact of the corrections because of their normalization to the LO
distribution. Normalizing the corrections to
the QCD-corrected differential cross section would reveal that the
impact of $\text{QCD}\times\text{EW}$ corrections is back to the few-percent level.
As for the applicability of the PA for $k_{\rT,\ell}>\MZ/2$, where the 
LO distribution receives only contributions from off-shell $\PZ$~bosons,
we still expect a good approximative quality of the PA, because the recoil
from QCD ISR effects, which is part of the $\text{QCD}\times\text{EW}$ corrections, 
allows for the population of this phase-space region by resonant $\PZ$~bosons.
Comparing our PA results on corrections to the $k_{\rT,\ell}$
distribution shown in the left--middle plot of \reffi{fig:NNLOdist} to the
corresponding results of \citere{Bonciani:2021zzf} for bare muons, shown in their Fig.~1, 
we again find agreement within statistical fluctuations.

Finally, the $\text{QCD}\times\text{EW}$ corrections to the $y_{\ell\ell}$
distribution are about $0.5{-}1\%$ and, thus, phenomenologically less
important in the central region, where most of the events are concentrated.

\begin{figure}
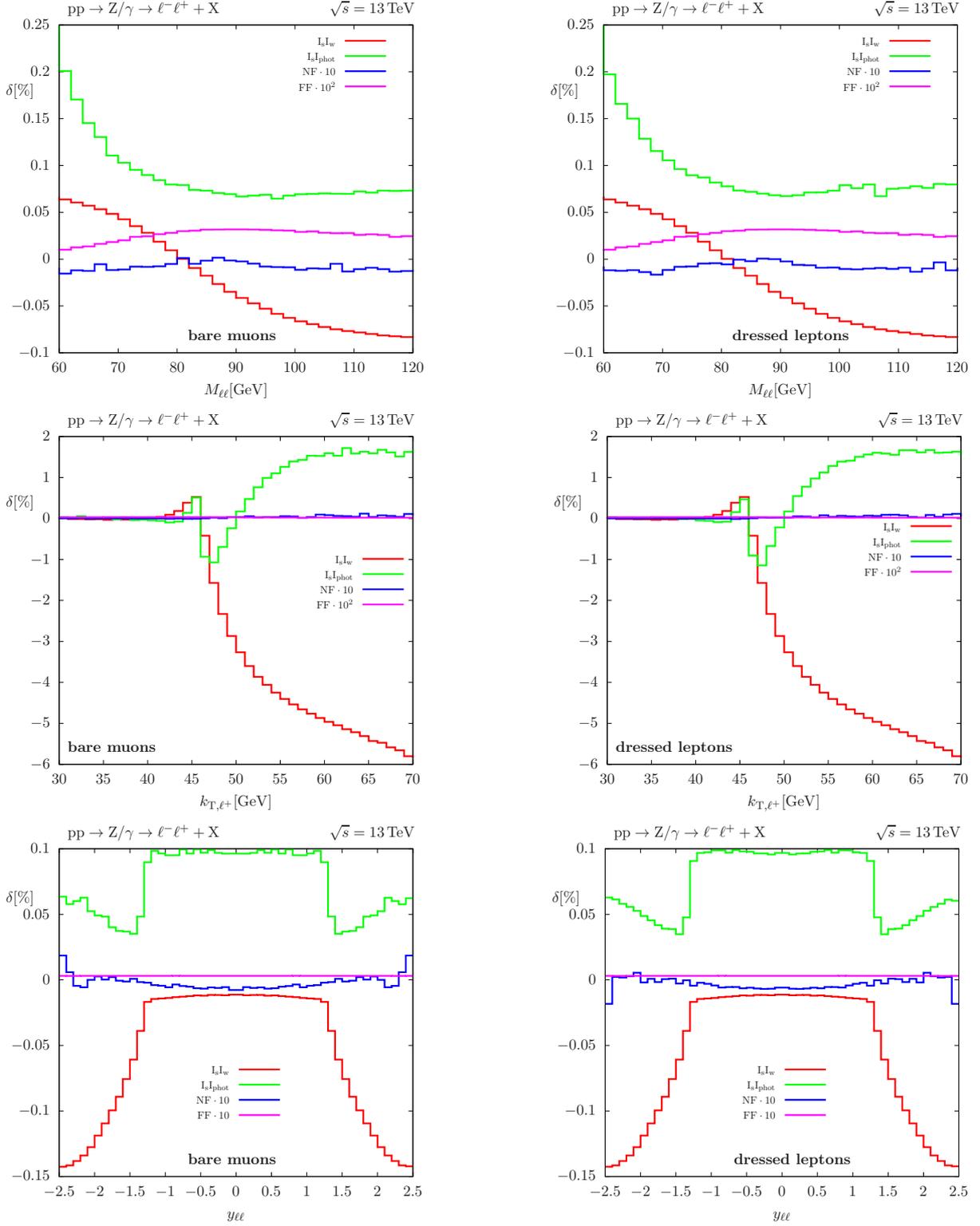

\includegraphics[scale=.65]{{{plots/NNLO_PA/Z-low.mll.NNLO_PA.bare}}}
\hfill
\includegraphics[scale=.65]{{{plots/NNLO_PA/Z-low.mll.NNLO_PA.dressed}}}
\includegraphics[scale=.65]{{{plots/NNLO_PA/Z-low.ktl1.NNLO_PA.bare}}}
\hfill
\includegraphics[scale=.65]{{{plots/NNLO_PA/Z-low.ktl1.NNLO_PA.dressed}}}
\includegraphics[scale=.65]{{{plots/NNLO_PA/Z-low.yll.NNLO_PA.bare}}}
\hfill
\includegraphics[scale=.65]{{{plots/NNLO_PA/Z-low.yll.NNLO_PA.dressed}}}
\caption{Subcontributions of the relative NNLO $\text{QCD}\times\text{EW}$ corrections in PA
(normalized to LO):
II $\text{QCD}\times\text{weak}$ (red),
II $\text{QCD}\times\text{photonic}$ (green),
NF (blue),
and FF $\text{QCD}\times\text{weak}$ (magenta).
The dominating IF contribution is contained in \reffi{fig:NNLOdist}.}
\label{fig:NNLOdistdecomp}
\end{figure}

The $\text{QCD}\times\text{EW}$ corrections other than the IF~contribution,
i.e.\ the corrections of
types~II (calculated in this paper), 
NF~\cite{Dittmaier:2014qza}, and FF~\cite{Dittmaier:2015rxo} 
are depicted in \reffi{fig:NNLOdistdecomp}. 
\begin{figure}
\includegraphics[scale=.65]{{{plots/NNLO_photos/Z-low.mll.NNLO.IF.photos.bare}}}
\hfill
\includegraphics[scale=.65]{{{plots/NNLO_photos/Z-low.mll.NNLO.IF.photos.dressed}}}
\includegraphics[scale=.65]{{{plots/NNLO_photos/Z-low.ktl1.NNLO.IF.photos.bare}}}
\hfill
\includegraphics[scale=.65]{{{plots/NNLO_photos/Z-low.ktl1.NNLO.IF.photos.dressed}}}
\includegraphics[scale=.65]{{{plots/NNLO_photos/Z-low.yll.NNLO.IF.photos.bare}}}
\hfill
\includegraphics[scale=.65]{{{plots/NNLO_photos/Z-low.yll.NNLO.IF.photos.dressed}}}
\caption{NNLO IF $\text{QCD}\times \text{EW}$ (green) and $\text{dNLO}_\text{QCD}\otimes\text{dPhotos}$ (red) corrections, normalized to LO.}
\label{fig:NNLO-photos}
\end{figure}
For the shown window around the $\PZ$~resonance in the $M_{\ell\ell}$ distribution and
in the whole $y_{\ell\ell}$ distribution, these contributions never exceed $0.2\%$
and are phenomenologically not relevant in the absolute predictions of the 
differential cross sections.
Only the II~corrections to the $k_{\rT,\ell}$ distribution gain some relevance, reaching
the $1\%$~level near $k_{\rT,\ell}\sim\MZ/2$. Recall again that for larger
$k_{\rT,\ell}$ the corrections should be normalized to the QCD-corrected prediction for an
assessment of their true impact.

Figure~\ref{fig:NNLO-photos} compares the IF factorizable corrections with an approximation obtained by folding the NLO QCD \emph{correction}, $\mathrm{dNLO}_\text{QCD}$, with the \textsc{Photos} QED shower.
In this way, terms of $\mathcal{O}(\alphas\alpha)$ are generated that can be contrasted with the PA calculation.
Overall, we observe a good qualitative agreement between the two predictions, further supporting our observation that the dominant effects at this order arise from FSR QED effects.
Moreover, the good agreement hints that the impact of multi-photon emissions beyond what is captured in our calculation is likely not of high phenomenological relevance.
As we will see in the next section, the agreement seen here in the absolute predictions degrade visibly in the case of the forward--backward asymmetry $A_\text{FB}$.

\subsection{Corrections to the Forward--backward asymmetry}
The forward--backward (FB) asymmetry 
for $\ell^+\ell^-$ production at the LHC is defined as \cite{Baur:1997wa,Baur:2001ze}
\begin{align}
 A_\text{FB}(M_{\ell\ell})=\frac{\sigma_\rF(M_{\ell\ell})-\sigma_\rB(M_{\ell\ell})}{\sigma_\rF(M_{\ell\ell})+\sigma_\rB(M_{\ell\ell})}
 \label{eq:def-Afb}
\end{align}
with
\begin{align}
\sigma_\rF(M_{\ell\ell}) = \int_0^1 \text{d}\!\cos \theta^*\, 
\frac{\text d \sigma}{\text{d}\!\cos \theta ^*}, & & 
\sigma_\rB(M_{\ell\ell}) = \int_{-1}^0 \text{d}\!\cos \theta^*\, 
\frac{\text d \sigma}{\text{d}\!\cos \theta ^*}.
 \label{eq:def-Af-Ab}
\end{align}
The angle $\theta^*$ is the so-called Collins--Soper (CS) angle, which is defined 
by~\cite{Collins:1977iv,Baur:1997wa}
\begin{align}
\cos \theta^* = \frac{\vert k^3_{\ell\ell}\vert}{k^3_{\ell\ell}} 
\frac{2}{M_{\ell\ell}\sqrt{M_{\ell\ell}^2+k_{\rT,\ell\ell}^2}} 
\Big(k^+_\ell k^-_{\bar\ell}-k^+_{\bar\ell}k^-_\ell\Big),
 \label{eq:def-cs-angle}
\end{align}
where 
\begin{align}
k^\mu_{\ell\ell} = k^\mu_{\ell} + k^\mu_{\bar\ell}, \qquad
 k^\pm_j = \frac{1}{\sqrt{2}} (k^0_j\pm k^3_j), \quad j=\ell,\bar\ell,
\end{align}
i.e.\ $M_{\ell\ell}^2=k_{\ell\ell}^2$.
All four-momenta are defined in the LAB frame.

The FB asymmetry $A_\text{FB}(M_{\ell\ell})$
is mainly relevant for determining the leptonic effective weak mixing angle
$\sin^2\theta_{\mathrm{w,eff}}^\ell$
defined on the $\PZ$~resonance from an $M_{\ell\ell}$ window around $\MZ$ with a width of
about $\sim\pm10\GeV$. 
LEP/SLC precision in $\sin^2\theta_{\mathrm{w,eff}}^\ell$ roughly translates 
into an uncertainty of $\sim10^{-3}$ in $A_\text{FB}$, so that the precision target
for an improved determination of $\sin^2\theta_{\mathrm{w,eff}}^\ell$ at the
LHC requires to control the prediction of $A_\text{FB}(M_{\ell\ell})$ at the
level of few $\sim10^{-4}$ in the vicinity of the $\PZ$~resonance.
Existing measurements of $A_\text{FB}$ at the LHC~\cite{ATLAS:2015ihy,LHCb:2015jyu,CMS:2016bil,CMS:2018ktx}
are already at the accuracy level of $10^{-3}$ near the $\PZ$~resonance.
Increased statistics from higher luminosity and steady improvements in the determination of parton distribution functions will tackle two of the main sources of uncertainties in these measurements, further challenging the precision of the underlying theory predictions.

\begin{figure}
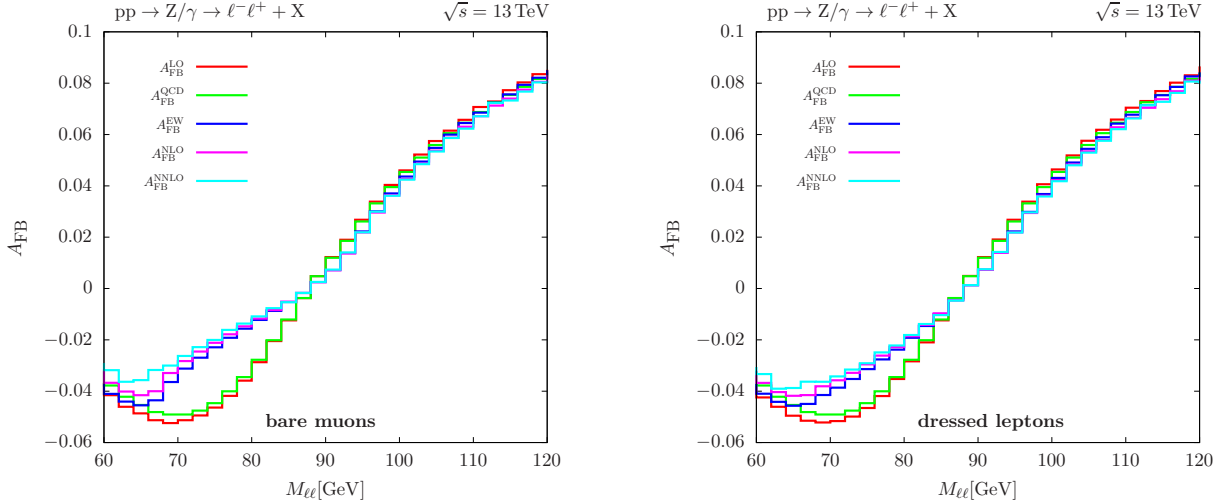

\centering
\includegraphics[scale=.65]{{{plots/NNLO/Z-Afb.NLO+NNLO.bare}}}
\hfill
\includegraphics[scale=.65]{{{plots/NNLO/Z-Afb.NLO+NNLO.dressed}}}
\caption{FB asymmetry $A_\text{FB}$ for muon pair (left)
and dressed-lepton pair (right) production at LO (red) 
and including various corrections:
NLO QCD (green), NLO EW (blue), full NLO QCD + EW (pink), 
and NNLO $\equiv$ NLO + EWHO + $\Delta$LLFSR + QCD$\times$QCD + QCD$\times$EW (light blue).}
\label{fig:Afb-LO}
\end{figure}

In \reffi{fig:Afb-LO} the LO prediction for the FB asymmetry $A_\text{FB}$ 
is compared to predictions including NLO QCD, NLO EW, and 
all available NNLO corrections, where
\begin{align}
 A_\text{FB}^x &= \frac{\sigma^x_\rF-\sigma^x_\rB}{\sigma^x_\rF+\sigma^x_\rB},
\qquad x=\LO,\, \NLO,\, \NNLO,
  \label{eq:def-Afbx}
\end{align}
where $\sigma^x_{\rF/\rB}$ are the forward/backward cross sections \refeq{eq:def-Af-Ab}
evaluated at order~$x$. 
The absolute prediction for $A_\text{FB}(M_{\ell\ell})$, which is shown in \reffi{fig:Afb-LO}, 
is of the order of
$10^{-2}$ near the $\PZ$~resonance, and the impact of NLO and NNLO corrections 
is already visible there.
To quantify the impact of the various corrections better,
we consider the shifts with respect to the LO asymmetry:
\begin{align}
  \Delta A_\text{FB}^{x} ={}& A_\text{FB}^{\text{LO}+\delta^x}-A_\text{FB}^\text{LO}, 
 \label{eq:def-DAfb}
\end{align}
where $\delta^x$ indicates the higher-order correction of type~$x$.

Figure~\ref{fig:dAfb-LO-NLO} separately shows the impact of NLO QCD, NLO EW, $\gamma$-induced,
and the pure photonic FSR part of the NLO EW corrections.
\begin{figure}
\centering
\includegraphics[scale=.65]{{{plots/NLO/Z-dAfb.NLO.bare}}}
\hfill
\includegraphics[scale=.65]{{{plots/NLO/Z-dAfb.NLO.dressed}}}
\caption{NLO corrections to the FB asymmetry $A_\text{FB}$ for muon pair (left)
and dressed-lepton pair (right) production:
NLO QCD (black), NLO EW (red), $\gamma$-induced contribution (cyan),
full NLO QCD + EW (green), and
photonic FSR at NLO (blue).}
\label{fig:dAfb-LO-NLO}
%\end{figure}
\vspace{3cm}
%
%\begin{figure}
\includegraphics[scale=.65]{{{plots/NLO_EW/Z-dAfb.NLO_EW.bare}}}
\hfill
\includegraphics[scale=.65]{{{plots/NLO_EW/Z-dAfb.NLO_EW.dressed}}}
\caption{NLO EW corrections to the FB asymmetry $A_\text{FB}$ for muon pair (left)
and dressed-lepton pair (right) 
production induced by QED IF (red), QED ISR (green), and 
purely weak (blue) corrections,
as well as contributions from LO $\gamma\gamma$ (cyan)
and NLO $q\gamma/\bar q\gamma$ (yellow) initial states.}
\label{fig:dAfb-EW}
\end{figure}
The latter constitutes the dominating effect with an impact of $\sim10^{-2}$
at the edge of the $\MZ-10\GeV<M_{\ell\ell}<\MZ+10\GeV$ window around the $\PZ$~resonance.
NLO QCD corrections are in the ballpark of $\sim10^{-3}$, while the photon-induced contributions are strongly suppressed as expected.
The NLO EW corrections not connected to FSR are illustrated in 
\reffi{fig:dAfb-EW} together with the separation of the $\gamma$-induced contributions into LO and NLO parts.
After the FSR effects, the most prominent contribution at NLO EW is given by
the purely weak corrections, which reach up to $5\times10^{-3}$.
The photonic IF interference, photonic ISR, and $\gamma$-induced effects typically contribute only
fractions of $10^{-3}$, but have to be taken into account in predictions at the targeted
level of precision.
We note that the LO $\gamma\gamma$ contribution is symmetric in the forward and backward directions, however, the change in the (symmetric) denominator of Eq.~\eqref{eq:def-Afbx} gives rise to a non-vanishing effect in Eq.~\eqref{eq:def-DAfb} seen in the figures.

\begin{figure}
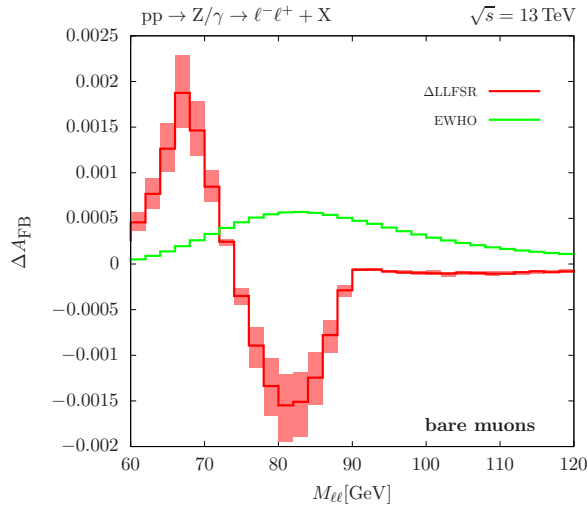

\centering
\includegraphics[scale=.65]{{{plots/LLFSR_EWHO/Z-dAfb.EWHO_LLFSR.bare}}}
\caption{NNLO $\Delta$LLFSR (red) and EWHO corrections to the forward--backward asymmetry. Within the red band the scale $\mu = \mu_\text{F}=\mu_\text{R}$ is varied in the calculation of the LLFSR corrections in the range $M_Z/2 < \mu < 2 M_Z$.}
\label{fig:dAfb-NNLO-EWHO}
\end{figure}

Figure~\ref{fig:dAfb-NNLO-EWHO} shows the effect of multi-photon emission
off leptons and of the universal higher-order EW corrections beyond NLO on the
FB asymmetry.
The higher-order FSR corrections modify $A_\text{FB}$ at the level of $\sim10^{-3}$ with
a residual scale uncertainty of a few $10^{-4}$ and are, thus, important.
On the other hand, the NNLO universal EW corrections amount only to a few $10^{-4}$.

\begin{figure}
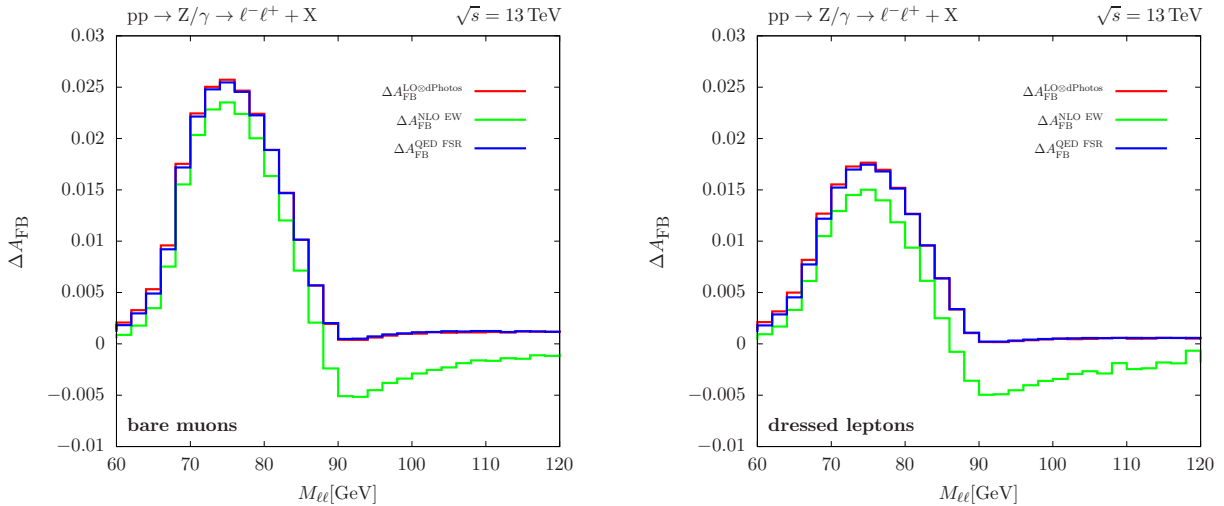

\centering
\includegraphics[scale=.65]{{{plots/NLO_photos/Z-dAfb.NLO.photos.bare}}}
\hfill
\includegraphics[scale=.65]{{{plots/NLO_photos/Z-dAfb.NLO.photos.dressed}}}
\caption{NLO $\text{EW}$ (green) and $\text{LO}\otimes\text{dPhotos}$ (red) corrections to the FB asymmetry.}
\label{fig:dAfb-NLO-photos}
\end{figure}%
In \reffi{fig:dAfb-NLO-photos}
we contrast full NLO EW corrections and QED FSR effects with a prediction based on the \textsc{Photos} QED shower on top
of the LO prediction.
We see good agreement between FSR corrections and the QED shower and the small 
normalization difference that was observed in the absolute predictions of 
\reffi{fig:NLO-photos} largely drop out in the observable $A_\text{FB}$.
A notable difference to the absolute predictions discussed in the previous section 
is the much more pronounced impact of non-FSR effects as can be assessed from 
a comparison to the full NLO EW curve.
This does not come as a surprise given the strong sensitivity of this observable to 
the weak sector of the Standard Model.

In \reffi{fig:dAfb-NNLO} we show the impact of the NNLO QCD and 
$\text{QCD}\times \text{EW}$ corrections 
in PA, together with the dominating
IF~contribution of the latter.
\begin{figure}
\centering
\includegraphics[scale=.65]{{{plots/NNLO/Z-dAfb.NNLO.bare}}}
\hfill
\includegraphics[scale=.65]{{{plots/NNLO/Z-dAfb.NNLO.dressed}}}
\caption{NNLO $\text{QCD}\times \text{QCD}$ corrections (red),
full $\text{QCD}\times \text{EW}$ corrections in PA (blue), 
and IF $\text{QCD}\times \text{EW}$ corrections in PA (green) to the FB asymmetry.}
\label{fig:dAfb-NNLO}
%\end{figure}
%
\vspace{3cm}
%\begin{figure}
\centering
\includegraphics[scale=.65]{{{plots/NNLO_PA/Z-dAfb.NNLO_PA.bare}}}
\hfill
\includegraphics[scale=.65]{{{plots/NNLO_PA/Z-dAfb.NNLO_PA.dressed}}}
\caption{NNLO $\text{QCD}\times \text{EW}$ corrections of types FF (red), NF (green), 
I$_s$I$_\text{p}$ (blue), and I$_s$I$_\text{w}$ (purple) to the FB asymmetry.}
\label{fig:dAfb-NNLO-QCDEW}
\end{figure}%
\begin{figure}
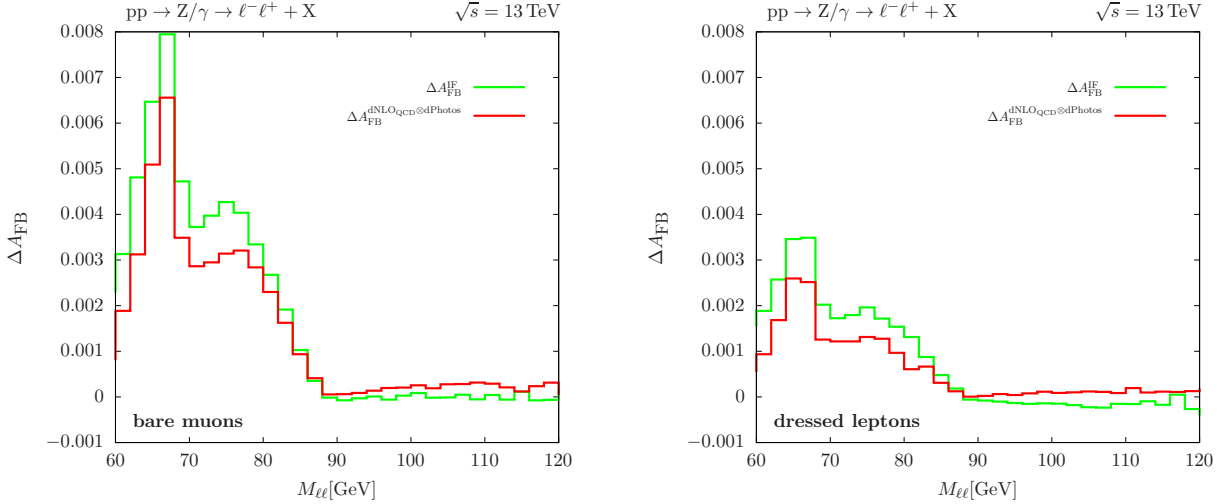

\centering
\includegraphics[scale=.65]{{{plots/NNLO_photos/Z-dAfb.NNLO.IF.photos.bare}}}
\hfill
\includegraphics[scale=.65]{{{plots/NNLO_photos/Z-dAfb.NNLO.IF.photos.dressed}}}
\caption{NNLO IF $\text{QCD}\times \text{EW}$ (green) and $\text{dNLO}_\text{QCD}\otimes\text{dPhotos}$ (red) corrections to the FB asymmetry.}
\label{fig:dAfb-NNLO-photos}
\end{figure}%
The $\text{QCD}\times \text{EW}$ IF~corrections change the FB asymmetry in the
vicinity of the $\PZ$~resonance by a few $10^{-3}$ and are, thus, phenomenologically
very important, while the NNLO QCD corrections contribute only at the level of
few $10^{-4}$.
The $\text{QCD}\times \text{EW}$ corrections of types other than IF are 
depicted in \reffi{fig:dAfb-NNLO-QCDEW}.
While the corrections of type~FF still reach the relevant level of $4\times10^{-4}$,
the $\text{QCD}\times \text{photonic}$ and $\text{QCD}\times \text{weak}$
II~corrections as well as the NF contributions are well below $10^{-4}$
and, thus, phenonemonologically negligible.

Finally, Figure~\ref{fig:dAfb-NNLO-photos} shows the comparison of 
the results obtained with the QED shower \textsc{Photos} on top of the NLO QCD prediction,
$\mathrm{dNLO}_\text{QCD}\otimes\textsc{Photos}$, with the IF factorizable 
$\mathcal{O}(\alphas\alpha)$ corrections.
A similar picture emerges here as at the previous order where the agreement between
the two results 
that was seen in the absolute predictions of \reffi{fig:NNLO-photos} 
substantially degrades in this observable, with a notable shape distortion that pivots 
around the resonance.
This is again likely due to the more sizeable impact of the non-FSR 
contributions that are not captured by a QED shower.

\section{Summary}
\label{se:summary}

Next-to-next-to-leading-order (NNLO) corrections of mixed $\text{QCD}\times \text{electroweak (EW)}$
origin, together with the recent process in third-order ($\textnormal{N}^3\textnormal{LO}$) QCD results,
are among the most important
fixed-order corrections beyond the well-known NNLO QCD and 
next-to-leading-order (NLO) EW corrections to differential cross sections of
Drell--Yan-like lepton pair production. 
In the vicinity of the $\PZ$-boson resonance,
the pole approximation (PA) can be used to reduce the complexity of the NNLO
$\text{QCD}\times \text{EW}$ corrections significantly.
The PA allows to classify the corrections into four separately
gauge-invariant building blocks: corrections of types initial--initial (II), initial--final (IF), 
final--final (FF), and non-factorizable (NF) corrections.
Making use of previous calculations of the IF, FF, and NF corrections, in this paper
we have completed the PA at the order $\O(\alphas\alpha)$ by calculating the 
corrections of type~II.

Technically, we have split the $\O(\alphas\alpha)$ II~corrections of the PA into two
separately gauge-invariant parts:
the $\text{QCD}\times \text{photonic}$ corrections with photon exchange between
and photon radiation off quarks, which we have evaluated without using the PA,
and the $\text{QCD}\times \text{weak}$ corrections with additional $\PW/\PZ$ exchange
in loops in PA.
For the latter, we have recalculated the needed two-loop $\PZ\bar\Pq\Pq$ formfactor
and presented explicit analytical results. 
For the QCD infrared (IR) singularities, which
are only of NLO complexity in this part, we have applied antenna and
dipole subtraction. 
The IR singularities of the $\text{QCD}\times \text{photonic}$ corrections,
which are of NNLO complexity, are treated with antenna subtraction.
All our results have been derived in two completely independent calculations,
the results of which are in good mutual numerical agreement.

Although the $\O(\alphas\alpha)$ corrections to the full off-shell lepton pair
production process have been calculated in recent years, the completion of the PA is still
very useful.
Firstly, a detailed numerical comparison between the full calculation and the PA
sheds light on the structure of the $\O(\alphas\alpha)$ corrections, which might
be helpful in calculating or approximating such corrections for related processes.
The completion of the PA presented here renders such a comparison possible.
Secondly, the full off-shell calculation is extremely complex and numerically 
challenging. For this reason, a detailed discussion of 
the $\O(\alphas\alpha)$ corrections to the forward--backward asymmetry $A_\text{FB}$ of the
leptons that is fully differential wrt.\ the invariant mass of the lepton pair
was still missing in the literature. With the numerical results presented in this
paper we have closed this gap.

The prospects to measure the leptonic effective weak mixing angle in the
high-lu\-mi\-no\-si\-ty phase of the LHC with a precision exceeding LEP accuracy,
translates into a target precision in the predictions for $A_\text{FB}$ of
a few $10^{-4}$ in the $\PZ$~resonance region.
We have presented a detailed survey of higher-order corrections, 
comprising results at the NLO QCD + EW level,
the NNLO QCD + $\text{QCD}\times \text{EW}$ level, and leading EW effects
beyond NLO from multi-photon emission and universal EW corrections.
Photonic final-state radiation (FSR) at NLO produces the largest correction to
$A_\text{FB}$ of about $10^{-2}$, followed by the NLO weak corrections 
of about $5\times10^{-3}$. The remaining NLO contributions, including QCD,
affect $A_\text{FB}$ at the level of few $10^{-3}$.
The NNLO $\O(\alphas\alpha)$ corrections of IF~type, which combine
QCD corrections to $\PZ$~production and photonic FSR off the leptons modify
$A_\text{FB}$ at the level of few $1{-}2\times10^{-3}$, which is also the
typical size of multi-photon effects.
Finally, the NNLO QCD corrections and the remaining $\O(\alphas\alpha)$ corrections
generically matter at the level of a few $10^{-4}$.

As already mentioned,
one of the natural next steps in the evaluation of $\O(\alphas\alpha)$ corrections to
Drell--Yan-like processes is a detailed comparison of PA-based and full
off-shell results for all individual ingredients, such as photonic and weak corrections
to lepton pair production.
Last but not least, another important step will be the completion of both the PA and
the full off-shell calculation of $\O(\alphas\alpha)$ corrections to the
charged-current process of $\PW$~production, which will be particularly important for
upcoming high-precision measurements of the $\PW$-boson mass at the LHC.

\section*{Acknowledgements}

We thank Timo Schmidt for contributions in the calculation of integrals
in an early stage of this work.
SD and JS
acknowledge support by the state of Baden-W\"urttemberg through bwHPC
and the German Research Foundation (DFG) through grants no.\ INST 39/963-1 FUGG,
grant DI~785/1, and the DFG Research Training Group RTG2044.

\clearpage

\appendix
\section*{Appendix}

\section{\boldmath{Formfactors for the irreducible {$\mathcal{O}(\alphas\alpha_\Pw)$}
corrections to the {$\PZ \bar ff$} vertex}}
\label{se:app:formfactors}

Here we present our explicit analytical results for the
formfactor functions $\phi_{\mathrm{A}}(z)$ and
$\phi_{\mathrm{NA}}(z)$ defined in \citere{Kotikov:2007vr} to described the
irreducible contributions of ${\cal O}(\alphas\alpha_\Pw)$
to the $\PZ \bar ff$ vertex corrections, as defined in Eq.~\refeq{eq:FZff}.
These contributions correspond to the difference between the full
formfactors and the naive product of the corresponding 
of ${\cal O}(\alphas)$ QCD and ${\cal O}(\alpha_\Pw)$ weak contributions.
We prefer to call those parts ``irreducible'' rather than ``non-factorizable''
(as done in \citere{Kotikov:2007vr}) to avoid confusion with our
classification of the different types of corrections arising in the resonance expansion.

We have performed a completely independent recalculation of the formfactors 
employing standard methods.
In detail, we have generated 
the two-loop graphs with {\sc FeynArts}~1.0~\cite{Kublbeck:1990xc} and
further algebraically processed the amplitudes with inhouse {\sc Mathematica} routines, 
to express them in terms of scalar two-loop integrals. 
The large set of two-loop integrals is reduced to a set of master integrals
with Laporta's algorithm~\cite{Laporta:2001dd}, employing 
the program {\sc Kira}~2.0~\cite{Maierhoefer:2017hyi,Klappert:2020nbg}.
Finally, the master integrals are calculated via differential equations
using Henn's canonical $\epsilon$-form~\cite{Henn:2013pwa,Henn:2014qga},
directly producing a result in terms of Goncharov Polylogarithms 
(GPLs)~\cite{Goncharov:1998kja,Goncharov:2010jf}.

For the formfactor functions $\phi_{\mathrm{A}}(z)$ and
$\phi_{\mathrm{NA}}(z)$ of \citere{Kotikov:2007vr} we explicitly get
\begin{align}
\phi_{\mathrm{A}}(z) ={}&
 \frac{4(1 + z)^2}{3 z^2} \Bigl[
- 24 G(0, 1, 0, -1; -z) + 6 G(0, 1, 0, 0; -z) - 6 G(1, 0, 0, -1; -z) 
\nn\\
& \qquad {} 
- 12 G(1, 0, 1, 0; -z) + 36 G(1, 1, 0, -1; -z) - 12 G(1, 1, 0, 0; -z) 
\nn\\
& \qquad {} 
+ 12 G(1, 1, 1, 0; -z) - 18 G(1, 0, -1; -z) + 9 G(1, 0, 0; -z) 
\nn\\
& \qquad {} 
- 9 G(1, 1, 0; -z) + 2 \pi^2 G(0, 1; -z) - 5 \pi^2 G(1, 1; -z) 
\nn\\
& \qquad {} 
- 3 (6 \zeta_3-\pi^2 ) G(1; -z) 
\Bigr]
\nn\\
& {}
+ \frac{16 (1 + 3 z + z^2)}{3 z^2}  \Bigl[
- 3 G(0, -1, 0, -1; -z) + 3 G(0, -1, 0, 0; -z) 
\nn\\
& \qquad {}
+\pi^2 G(0, -1; -z) \Bigr]
\nn\\
& {}
+ \frac{12 (1 - z^2)}{z^2}  \Bigl[\pi^2 G(-1; -z) - 3 G(-1, 0, -1; -z) + 3 G(-1, 0, 0; -z)\Bigr]
\nn\\
& {}
+ \frac{(2 + 3 z)}{3 z} \Bigl[12 G(0, 1, 0; -z) + 36 \zeta_3+ 16 \pi^2 -51 G(0; -z) \Bigr]
\nn\\
& {}
+ \frac{4 (4 + 3 z)}{z}  G(0, 0, -1; -z)
- \frac{4 (11 + 9 z)}{z}  G(0, -1; -z)
\nn\\
& {}
+ \frac{2 (16 + 23 z)}{z}  G(0, 0; -z)
- \frac{2 (1 + z) (3 + 5 z)}{z^2}  G(1, 0; -z)
\nn\\
& {}
+ \frac{2 (-1 + z) (13 + 27 z)}{z^2}  G(-1; -z)
-2 + \frac{16}{z},
\\[.5em]
\phi_{\mathrm{NA}}(z) ={}&
\frac{8 (1 + z)^2}{3 z^2} \Bigl[ 
6 G(0, -1, 0, -1; -z) - 6 G(0, -1, 0, 0; -z) + 24 G(0, 1, 0, -1; -z) 
\nn\\
& \qquad {}
- 6 G(0, 1, 0, 0; -z) + 6 G(1, 0, 0, -1; -z) + 12 G(1, 0, 1, 0; -z) 
\nn\\
& \qquad {}
- 36 G(1, 1, 0, -1; -z) + 12 G(1, 1, 0, 0; -z) - 12 G(1, 1, 1, 0; -z) 
\nn\\
& \qquad {}
+ 18\zeta_3  G(1; -z) + 5 \pi^2 G(1, 1; -z) 
+ 18 G(1, 0, -1; -z) - 9 G(1, 0, 0; -z) 
\nn\\
& \qquad {}
+ 9 G(1, 1, 0; -z) 
-3 \pi^2 G(1; -z) - 2 \pi^2 G(0, -1; -z) - 2 \pi^2 G(0, 1; -z) 
\Bigr]
\nn\\
& {}
+ \frac{8 (1- z^2)}{z^2}  \Bigl[ 3 G(-1, 0, -1; -z) - 3 G(-1, 0, 0; -z) -\pi^2 G(-1; -z) \Bigr]
\nn\\
& {}
- \frac{8 (4 + 3 z)}{z}  G(0, 0, -1; -z)
- \frac{8 (2 + 3 z)}{z}  G(0, 1, 0; -z)
\nn\\
& {}
- \frac{4 (1 + 3 z) (5 + 3 z)}{z^2}  G(0, -1; -z)
+ \frac{4 (1 + z) (3 + 5 z)}{ z^2}  G(1, 0; -z)
\nn\\
& {}
- \frac{4 (1 - z) (13 + 19 z)}{z^2}  G(-1; -z)
- \frac{6 (2 + 7 z)}{z}  G(0; -z)
\nn\\
& {}
- \frac{48 (1 - y^2) (1 - 2 y) }{y^4}
   \Bigl[2 G(0, 1, 1, 1; y) - 2 G(0, 1, 1, y_2; y) - 2 G(0, 1, 1, y_1; y) 
\nn\\
& \qquad {}
- G(1, 1, 1, 1; y) + G(1, 1, 1, y_2; y) + G(1, 1, 1, y_1; y) \Bigr]
\nn\\
& {}
- 16 G(0, 0, 1; y) + 16 G(0, 0, y_2; y) + 16 G(0, 0, y_1; y) + 24 G(0, 1, 1; y) 
\nn\\
& {}
- 8 G(0, 1, y_2; y) - 8 G(0, 1, y_1; y) + 8 G(1, 0, 1; y) - 8 G(1, 0, y_2; y) 
\nn\\
& {}
- 8 G(1, 0, y_1; y) - 8 G(1, 1, y_2; y) - 8 G(1, 1, y_1; y) 
\nn\\
& {}
- \frac{4 (10 - 20 y - 4 y^2 + 14 y^3 + 5 y^4)}{y^4}  \Bigl[G(0, 1; y) - G(0, y_2; y) - G(0, y_1; y)\Bigr]
\nn\\
& {}
- \frac{4 (-11 + 34 y - 20 y^2 - 12 y^3 + 5 y^4)}{ y^4}  G(1, 1; y)
\nn\\
& {}
+ \frac{4 (-5 + 22 y - 16 y^2 - 10 y^3 + 2 y^4)}{y^4}  \Bigl[G(1, y_2; y) + G(1, y_1; y)\Bigr]
\nn\\
& {}
+\frac{2 (-26 + 64 y - 46 y^2 - 6 y^3 + 31 y^4)}{y^4}  G(1; y)
\nn\\
& {}
- \frac{4 (1 - y + y^2) (-13 + 13 y + 12 y^2)}{y^4}  \Bigl[G(y_2; y) + G(y_1; y)\Bigr]
\nn\\
& {}
- \frac{48 \zeta_3}{z}
- 72 \zeta_3
- \frac{4 \pi^2}{3} 
- \frac{60}{y^2} + \frac{60}{y} - \frac{60}{z} -16,
\end{align}
where
\begin{align}
z ={}& \frac{q^2}{M^2}+\ri 0, 
\qquad
y_{1,2} = \frac{1}{2}\bigl(1\pm\ri\sqrt{3}\bigr),
\nn\\[.5em]
y ={}& y(z) =\left\{ \begin{array}{lll}
{} [z-\sqrt{z(z-4)}]/2 & \mbox{ for } & q^2<0, \\
2z/[z+\sqrt{z(z-4)}] & \mbox{ for } & q^2>4M^2, \\
{} [z-\ri\sqrt{z(4-z)}]/2 & \rlap{ otherwise.} &
\end{array}\right.
\end{align}
The variable $y$ is a solution of the quadratic equation $0 = z(1-y) + y^2$
with the special property that the contour 
from $y(0)=0$ to $y(z)$
defined by $z=r+\ri 0$, $r\in\mathbb{R}$, is homotopic to the straight line
from $0$ to $y(r+\ri 0)$ in the complex $y$ plane going between the
singularities at $y_{1,2}$ without hitting them. 
This property guarantees the correct analytical 
behaviour when integrating the differential equations in the kinematical
variable $y$ via straight lines as contours in the complex $y$~plane
with $y=0$ ($q^2=0$) as start condition.
This directly leads to GPLs in $y$, without picking up
contributions from residues resulting from contour deformations.
Note that $y(1)=y_2$ lies on the curve $y(r)$
with $r\in\mathbb{R}$ which is obtained without taking Feynman's
$\ri0$ prescription into account, and the $\ri0$ prescription circumvents
$y_2$ in the ``correct direction''. A similar path defined from the second solution
of $0 = z(1-y) + y^2$ would circumvent $y_1$ in the ``wrong direction'',
so that a path defined analogously to the above $y(z)$ would not be 
homotopic to the straight line from $0$ to $y(r+\ri 0)$.

Since a direct comparison of our analytical results for
$\phi_{\mathrm{A}}(z)$ and $\phi_{\mathrm{NA}}(z)$ to the ones
given in \citere{Kotikov:2007vr} seems too cumbersome, we have compared
those functions numerically for the benchmark points given in 
\citere{Kotikov:2007vr}. The results of this comparison is shown
in \refta{tab:phiz}.
\begin{table}
\centerline{
\begin{tabular}{|l|l|l|}
\hline
& \multicolumn{1}{c|}{our result} & \multicolumn{1}{c|}{\citere{Kotikov:2007vr}}
\\
\hline
$\phi_{\mathrm{A}}(1+\ri 0)$  & $-2.1073169              -19.033126         \ri$ &
%$\phi_{\mathrm{A}}(1+\ri 0)$ & $-2.1073168595023617     -19.033126361554203\ri$ &
%                                -2.1073168582811164e+00,-1.9033126361352885e+01)
$-2.1073 - 19.0331\ri$
\\
$\phi_{\mathrm{NA}}(1+\ri 0)$  & $-7.5879998             +16.719365         \ri$ &
%$\phi_{\mathrm{NA}}(1+\ri 0)$ & $-7.5879997688372498    +16.719364983919537\ri$ &
%                                 -7.5879998762524670e+00,1.6719364694996244e+01
$-7.5880 + 16.7194\ri$
\\
\hline
$\phi_{\mathrm{A}}(1.2856+\ri 0)$  & $-1.3610142              -22.238253         \ri$ &
%$\phi_{\mathrm{A}}(1.2856+\ri 0)$ & $-1.3610141529025270     -22.238253047921873\ri$ &
%                                     -1.3610141518470513e+00,-2.2238253047608652e+01
$-1.3598 - 30.4095\ri$
\\
$\phi_{\mathrm{NA}}(1.2856+\ri 0)$  & $-10.121177             +18.696957         \ri$ &
%$\phi_{\mathrm{NA}}(1.2856+\ri 0)$ & $-10.121177130873981    +18.696956625842191\ri$ &
%                                      -1.0121177131513331e+01,1.8696956624938196e+01
$-10.1248 + 35.0336\ri$
\\
\hline
\end{tabular} }
\caption{Comparision our numerical results on $\phi_{\mathrm{A}}(z)$ and 
$\phi_{\mathrm{NA}}(z)$ with the benchmark numbers given in \citere{Kotikov:2007vr}.}
\label{tab:phiz}
\end{table}
While our results confirm the ones given in \citere{Kotikov:2007vr}
for $z=1+\ri0$ for all digits given there, we find significant 
differences for $z=1.2856+\ri 0$; in particular, the imaginary
parts of $\phi_{\mathrm{A}}(1.2856+\ri 0)$ and 
$\phi_{\mathrm{NA}}(1.2856+\ri 0)$ are totally different in the two evaluations. 
For this reason, we have implemented and numerically evaluated the analytical results of
\citere{Kotikov:2007vr} as well. The corresponding third set of results
completely confirms our results given in \refta{tab:phiz} to all digits
given there.
We therefore conclude that the analytical results of \citere{Kotikov:2007vr}
are in fact in agreement with ours, but the benchmark numbers on
$\phi_{\mathrm{A}}(1.2856+\ri 0)$ and $\phi_{\mathrm{NA}}(1.2856+\ri 0)$
given there are not correct.

%%%%%%%%%%%%%%%%%%%%%%%%%%%%%%%%%%%%%%%%%%%%%%%%%%%%%%%%%%%%%%%%%%%%%%
% \bibliographystyle{h-physrev}
\bibliographystyle{JHEPmod}
\bibliography{Oaas-ZPA-AFB.rev}

\end{document}